\documentclass[fleqn,usenatbib]{mnras}
\DeclareMathAlphabet{\mathcal}{OMS}{cmsy}{m}{n}
\DeclareMathAlphabet\mathbfcal{OMS}{cmsy}{b}{n}

\usepackage[T1]{fontenc}
\usepackage{graphicx}	% Including figure files
\usepackage{amsmath}	% Advanced maths commands
\usepackage{amssymb}	% Extra maths symbols
\usepackage{bm}
\usepackage{ulem}
\usepackage[dvipsnames]{xcolor}
\usepackage{tablefootnote}

\usepackage{chngpage}
\usepackage{multirow}
\usepackage{tabularx}

\usepackage[utf8]{inputenc}

\newcommand{\planck}{\textit{Planck}}
\newcommand{\euc}{\textit{Euclid}}
\newcommand{\hi}{\textsc{Hi}}
\newcommand{\cosmosis}{\texttt{CosmoSIS}}
\newcommand{\class}{\texttt{CLASS}}
\newcommand{\angpow}{\texttt{AngPow}}

\newcommand{\Multinest}{\texttt{Multinest}}

\newcommand{\lcdm}{$\Lambda$CDM}

\newcommand{\tcl}[2]{$C_\ell^{\rm g}(z_{#1},z_{#2})$}
\newcommand{\tcijl}[2]{$C_\ell^{\rm g}(z_{#1},z_{#2})$}
\newcommand{\de}{\mathrm{d}}
\newcommand{\ho}{H_0}
\newcommand{\om}{\Omega_{\rm m}}
\newcommand{\ob}{\Omega_{\rm b}}
\newcommand{\ns}{n_{\rm s}}
\newcommand{\As}{A_{\rm s}}
\newcommand{\lm}{$\ell_{\rm min}$}
\newcommand{\lM}{$\ell_{\rm max}$}
\newcommand{\Q}{\mathcal Q}
\newcommand{\T}{\mathcal T}
\newcommand{\Pk}{\mathcal P}

\title[LSS pipeline II: Magn.bias and radio continuum surveys]{Developing a unified pipeline for large-scale structure data analysis with angular power spectra -- II. A case study for magnification bias and radio continuum surveys}
\author[K.\ Tanidis,  S.\ Camera,  \& D.\ Parkinson]{
Konstantinos Tanidis,$^{1,2}$\thanks{E-mails: tanidis@to.infn.it (KT); stefano.camera@unito.it (SC); DavidParkinson@kasi.re.kr (DP).} Stefano Camera,$^{1,2,3}$ and David Parkinson$^4$
\\
$^{1}$Dipartimento di Fisica, Universit\`a degli Studi di Torino, via P.\ Giuria 1, 10125 Torino, Italy\\
$^{2}$INFN -- Istituto Nazionale di Fisica Nucleare, Sezione di Torino, via P.\ Giuria 1, 10125 Torino, Italy\\
$^{3}$Department of Physics \& Astronomy, University of the Western Cape, Cape Town 7535, South Africa\\
$^{4}$Korea Astronomy and Space Science Institute, 776 Daedeokdae-ro, Yuseong-gu, Daejeon 34055, Republic of Korea
}
\date{Accepted XXX. Received YYY; in original form ZZZ}
\pubyear{2019}

\begin{document}
\label{firstpage}
\pagerange{\pageref{firstpage}--\pageref{lastpage}}
\maketitle

\begin{abstract}
Following on our purpose of developing a unified pipeline for large-scale structure data analysis with angular power spectra, we now include the weak lensing effect of magnification bias on galaxy clustering in a publicly available, modular parameter estimation code. We thus forecast constraints on the parameters of the concordance cosmological model, dark energy, and modified gravity theories from galaxy clustering tomographic angular power spectra. We find that a correct modelling of magnification is crucial not to bias the parameter estimation, especially in the case of deep galaxy surveys. Our case study adopts specifications of the Evolutionary Map of the Universe, which is a full-sky, deep radio-continuum survey, expected to probe the Universe up to redshift $z\sim6$. We assume the Limber approximation, and include magnification bias on top of density fluctuations and redshift-space distortions. By restricting our analysis to the regime where the Limber approximation holds true, we significantly minimise the computational time needed, compared to that of the exact calculation. We also show that there is a trend for more biased parameter estimates from neglecting magnification when the redshift bins are very wide. We conclude that this result implies a strong dependence on the lensing contribution, which is an integrated effect and becomes dominant when wide redshift bins are considered. Finally, we note that instead of being considered a contaminant, magnification bias encodes important cosmological information, and its inclusion leads to an alleviation of the degeneracy between the galaxy bias and the amplitude normalisation of the matter fluctuations.
\end{abstract}

\begin{keywords}
cosmology: theory -- large-scale structure of the Universe -- observations -- cosmological parameters
\end{keywords}

%%%%%%%%%%%%%%%%%%%%%%%%%%%%%%%%%%%%%%%%%%%%%%%%%%

%%%%%%%%%%%%%%%%% BODY OF PAPER %%%%%%%%%%%%%%%%%%

\section{Introduction}
\label{sec:intro}
Our current understanding of the Universe's properties, evolution, and present-day composition has reached a degree of maturity unthinkable of only fifty years ago. This concordance picture tells us of an accelerating cosmic expansion at recent times---well accommodated by a cosmological constant, $\Lambda$---and of a large-scale structure (LSS) formed through accretion of inhomogeneities in the distribution of matter---mainly constituted by cold dark matter. This is the widely known \lcdm\ model, which has proven itself successful in describing the majority of the observations.

Undoubtedly, the cosmic microwave background (CMB) has hitherto been cosmology's treasure cove, and the \planck\ satellite final data release provided us with the tightest constraints on cosmological parameters \citep{2018arXiv180706209P}. However, most of the available information has been extracted by now---albeit the future of CMB studies is still bright, with the prospects of taming uncertainties on polarisation measurements down to the cosmic variance limit \citep[see e.g.][]{2016arXiv161002743A,Hazumi:2019lys,2019JCAP...02..056A}. Hence, there is nowadays a high level of expectations for LSS observational campaigns. Indeed, the LSS is potentially even more informative that the CMB, thanks to its three-dimensional nature (compared to the thin redshift slice around the last-scattering surface), and the study of the cosmic web can teach us about the nonlinear behaviour of gravity as well.

One of the main probes of the LSS is the clustering of galaxies, as it has been convincingly demonstrated in many an instance \citep[e.g.][]{2012PhRvD..86j3518P,2012MNRAS.423.3430B,2015MNRAS.449..848H,10.1093/mnras/stx721,2017A&A...604A..33P}. This field of research is nowadays entering a new era with the construction of a series of futuristic experiments. The forthcoming galaxy surveys will be game-changing probes of the LSS, observing from millions to billions of sources at different wavelengths and exploiting various techniques. A few examples of LSS experiments that will take data in the near future are: the European Space Agency's satellite \euc\ \citep{Laureijs2011,Amendola2013,Amendola2016}, the Large Synoptic Survey Telescope \citep{Abate:2012za}, or the Dark Energy Survey Instrument \citep{Aghamousa:2016zmz}, in the optican/near-IR band; and the Square Kilometre Array (SKA, \citealt{Maartens2015,Abdalla2015,SKA1_2018}) and its precursors, at radio frequencies.

For the aforementioned reasons, in a companion paper (\citealt{Tanidis:2019teo}; hereafter, Paper I) we set forth on a path to develop a unified pipeline for LSS data analysis with power spectra in harmonic space. We deem this a worthwhile purpose, urged by the consideration that the range of scales and redshifts probed by forthcoming surveys likely calls for a change of paradigm in the treatment of the data and the theoretical modelling. In the present paper, we focus on one of the SKA precursors, the Evolutionary Map of the Universe \citep[EMU,][]{norris2011} radio-continuum survey on the Australian SKA Pathfinder (ASKAP) telescope. Unlike the photometric (optical/near-IR) and the spectroscopic (optical/near-IR or \hi-line galaxy survey in the radio) experiments, radio continuum surveys like EMU have the advantage of being able to scan very quickly large areas of the sky by averaging over all frequencies, thus increasing the signal-to-noise ratio of each source. Despite the fact that the deep and fast scanning in redshift space can detect a large number of galaxies, including also very faint sources, their redshift estimation is quite poor. Given the insufficient redshift information, the angular tomographic clustering is usually adopted to analyse radio continuum galaxy catalogues, instead of the more usual three-dimensional Fourier-space power spectrum.

In this paper, we move past the Fisher matrix approach hitherto employed, to a full likelihood-based analysis. We particularly turn our interest to the investigation of the cosmological information encoded in the weak lensing effect of magnification bias on the density fluctuations of the galaxy field \citep[see][for a seminal review on gravitational lensing]{2001PhR...340..291B}. This effect is widely known and is due to the weak lensing contribution caused by the underlying matter field. It induces a modulation in the clustering signal across redshift bins, inducing a correlation between background and foreground sources.

%In this paper, however, we will investigate the effect of incorrectly neglecting the magnification bias for galaxy clustering alone on Bayesian inference, focusing on the case for radio continuum surveys with the EMU survey adopted as a proxy.
%
 The paper is outlined as follows. In \autoref{sec:Apower spectrum}, we introduce the harmonic-space angular power spectrum \tcijl{i}{j} with and without the magnification bias correction, and implement it in the publicly available \cosmosis\ code \citep{Zuntz2015}. In \autoref{sec:EMU_distribution}, we present the EMU survey specifications and simulation results used to construct the tomographic redshift bins that will be later applied in the analysis. In \autoref{sec:cuts}, we perform a comparison test between our Limber approximated \cosmosis\ code version and the full solution  obtained with \class\ \citep{Lesgourgues2011,Blas2011,DiDio2013}. In \autoref{sec:cosmo_theory}, we present the theoretical models considered, while the likelihood for the forecast is presented in \autoref{sec:like}. In \autoref{sec:results}, we examine in detail the Bayesian analysis of an idealistic and two realistic scenarios for the cosmological models considered, and we also show that the redshift-space distortions (RSDs) correction to the density field has negligible effect in our case. Finally, in \autoref{sec:conclusion}, we present our concluding remarks.

\section{Galaxy clustering in harmonic space}
\label{sec:Apower spectrum}

Here, we describe how to construct the galaxy clustering (tomographic) angular power spectrum, \tcl{i}{j}, including contributions from density fluctuations, RSDs, and magnification bias. To ensure the robustness of our cosmological results, we use only linear scales \citep[see][for a study on nonlinearities in angular spectra]{Jalilvand:2019brk} in a region where the Limber approximation holds true \citep{1953ApJ...117..134L,1992ApJ...388..272K}. In the following analysis, we implement this framework in a modified version of the \cosmosis\ package. The treatment in our analysis follows closely that of Paper I, to which we refer the reader for any clarification.

%\subsection{The matter power spectrum}
%\label{sec:Pk}
%
Let us start from the linear Fourier-space matter power spectrum,
\begin{equation}
P_{\rm lin}(k,z)=
\frac{8\pi^2}{25}\ho^{-4}\om^{-2}g_\infty^{-2}D^2(z)T^2(k)\Pk_\zeta(k)k,
%\nonumber\\&=P_{\rm lin}(k)D^2(z),
\label{eq:matter}
\end{equation}
where $\om$ is the total matter fraction in the Universe, and $\ho$ the Hubble constant at present. Furthermore, we take advantage of the fact that scale and redshift dependence can be considered separately when the anisotropic stress is not present, as in general relativity after radiation domination. Thus, a scale-dependent transfer function $T(k)$ and a redshift-dependent growth factor $D(z)$ can be defined, while $g_\infty=\lim_{z\to\infty}(1+z)D(z)\simeq1.27$. (Normalisations require $D(z)=1$ at $z=0$ and $T(k)=1$ for $k\to0$.) The term $\Pk_\zeta(k)=\As(k/k_0)^{\ns-1}$ is the dimensionless primordial curvature power spectrum, with $\As$ being the amplitude and $\ns$ the spectral index. Hereafter, we shall often use the shorthand notation $P_{\rm lin}(k)\equiv D^{-2}(z)P_{\rm lin}(k,z)=\T^2(k)\Pk_\zeta(k)$, which represents linear matter power spectrum at present; we also define $\T(k)=(4/5)\pi\ho^{-2}\om^{-1}g_\infty^{-1}T(k)k^{1/2}$.

\subsection{Galaxy number counts and magnification bias}
It is well known that light ray paths experience deflections by the intervening matter distribution lying on the line-of-sight direction. This induces distortions in the images of distant objects; such distortions, in the weak lensing limit, are usually decomposed into a `convergence' $\kappa$ and a `shear' $\gamma$. The former---a surface mass density integrated along the line of sight---is responsible for changing the apparent size of a distant galaxy's image, whilst the latter---a complex, spin-2 quantity---stretches an observed galaxy's shape in different directions, making for instance ellipses out of circles \citep[see][for some beautiful and intuitive illustrations of lensing distortions]{Clarkson:2016ccm}. In turn, convergence and shear are jointly responsible for the magnification,
\begin{equation}
    \mu=\left|(1-\kappa)^2-|\gamma|^2\right|^{-1}.
\end{equation}

Cosmic magnification has been first measured by cross-correlating high-redshift quasars with the low-redshift galaxies observed by the Sloan Digital Sky Survey \citep{Scranton_2005}, and later with galaxy-dust and galaxy-mass correlations by \cite{10.1111/j.1365-2966.2010.16486.x}. The same effect was detected with normal galaxy samples using the Canada-France-Hawaii-Telescope Legacy Survey measurements \citep{Hildebrandt}. Furthermore, the magnification bias has been proposed as a probe for the investigation of the primordial magnetic fields \citep{Camera_2014}.

Besides being a lensing observable per se \citep[e.g.][]{VanWaerbeke:2009fb}, magnification contributes to the observed correlation of galaxy number counts \citep{Yoo2010,ChallinorandLewis2011,BonvinDurrer2011}. The effect of magnification on the observed clustering is due to foreground galaxies effectively acting as lenses for sources in the background. On the one hand, images of a fixed set of sources are distributed over a larger solid angle, thus reducing the number density by a factor $\mu^{-1}$. On the other hand, the magnification allows for the observation of fainter sources, as the flux threshold is likewise lowered by the $\mu^{-1}$ factor. Now, if $N_{\rm g}$ is the \textit{comoving} number density of galaxies above a certain flux threshold $F^\ast$ (or, equivalently, below some magnitude threshold $m^\ast$), we can define,
\begin{align}
    \Q&=-\left.\frac{\partial\ln N_{\rm g}}{\partial\ln F}\right|_{F^\ast} \nonumber \\
    &=\frac52\left.\frac{\partial\log_{10} N_{\rm g}}{\partial m}\right|_{m^\ast}.\label{eq:magbias}
\end{align}
Hence, in the weak lensing regime where $\mu\approx 1+2\kappa$, it can be seen that the fluctuations in galaxy number counts, $\delta_{\rm g}$, get a further contribution from lensing. This is modulated by $\Q$, for which reason is called `magnification bias'.\footnote{An alternative notation is also known in the literature, with $s=2\Q/5$.} Specifically,
\begin{equation}
    \delta_{\rm g}=b\,\delta+\frac{(n^i\partial_i)^2}{\mathcal H}V+2(1-\Q)\kappa,\label{eq:delta_g}
\end{equation}
where $b$ is the linear galaxy bias, $\delta$ is the matter density contrast (expressed in the comoving-synchronous gauge), $\mathcal H$ is the conformal Hubble factor, $V$ is the velocity potential, $n^i$ denotes a galaxy's line-of-sight direction, and the calculation is performed in the longitudinal gauge. The first term in \autoref{eq:delta_g} is the usual Newtonian density fluctuations, the second term is RSDs, and the last is the magnification contribution.

The inclusion of the lensing magnification in cross and auto-correlations of galaxy clustering and cosmic shear has been studied with Fisher analysis \citep{10.1093/mnras/stt2060,Villa:2017yfg,Thiele:2019fcu,VanessaBohm}, where it has already been suggested that the ignorance of the magnification bias may induce a bias in the cosmological parameter estimation. Here, we test this hypothesis by performing a full likelihood mock data analysis.
%where $\delta_m$ is the dark matter distribution, $b$ the galaxy bias and $k$ the convergence defined as \citep{2001PhR...340..291B}:
%\begin{equation}
%    k(\pmb{\theta})=\int_0^{\chi_h} w^i(\chi)\delta_m(\pmb{\theta},\chi),
%\end{equation}
%and the weight function reads:
%\begin{equation}
%w^i(\chi)=\frac{3\om \ho^2}{2c^2}\frac{d_\theta (\chi)}{a(\chi)}\int_\chi^{\chi_h} \de\chi^\prime n^i(\chi^\prime)\frac{d_\theta(\chi-\chi^\prime)}{d_\theta(\chi^\prime)}    ,
%\end{equation}
%where $\chi_h$ is the comoving distance at the horizon, $a(\chi)$ the Universe scale factor at comoving distance $\chi$, $c$ the speed of light and $d_\theta$ the angular diameter distance.

\subsection{The observed galaxy number count angular power spectrum}
The galaxy number count angular spectrum on linear scales can be written as
\begin{equation}
C^{\rm g}_{\ell}(z_i,z_j)=4\upi\int\de\ln k\,\mathcal P_\zeta(k)\mathcal W_\ell^{\rm g}(k;z_i)\mathcal W_\ell^{\rm g}(k;z_j),
\label{eq:Cl_fullsky}
\end{equation}
with the redshift-integrated kernels given by
\begin{multline}
\mathcal W^{\rm g}_\ell(k;z_i)=\T(k)\int\de\chi\,D(\chi)\Big\{b(\chi)n^i(\chi)j_\ell(k\chi)\\
-f(\chi)n^i(\chi)j^{\prime\prime}_\ell(k\chi)+2\left[\Q(\chi)-1\right]w_\ell^{\kappa,i}(k,\chi)j_\ell(k\chi)\Big\},
\label{eq:weight_func}
\end{multline}
where $\chi$ is the comoving distance to redshift $z$,  $j_\ell$ the $\ell$th-order spherical Bessel function, $f\equiv-(1+z)\de\ln D/\de z$ is the growth rate,
\begin{equation}
w_\ell^{\kappa,i}(k,\chi)=\frac{3\om\ho^2}{2k^2}\left[1+z(\chi)\right]\ell(\ell+1)\tilde n^i(\chi)
\label{eq:convergence}
\end{equation}
is the lensing weight for the galaxy redshift distribution in the $i$th redshift bin, $n^i(\chi)$, and we have defined
\begin{equation}
\tilde n^i(\chi)=\int_\chi^\infty\de\chi^\prime\,\frac{\chi^\prime-\chi}{\chi^\prime\chi}n^i(\chi^\prime).
\end{equation}
Note that, unless otherwise stated, $\int\de z\,n^i(z)=1$ and $n^i(\chi)\de\chi=n^i(z)\de z$ hold true.

If we compare \autoref{eq:weight_func} to \autoref{eq:delta_g}, the effect of projecting in harmonic space becomes clear:
\begin{itemize}
\item Each different contribution to the fluctuations in the galaxy number density, $\delta_{\rm g}$, is modulated by a peculiar quantity---the bias for the matter density contrast, the growth rate for the RSDs, and the magnification bias for the lensing convergence.
\item Each contribution is weighted by the galaxy distribution in the redshift bin considered---note that lensing convergence is an integrated effect, weighted by a geometric factor, so that the source redshift distribution does not enter explicitly the third term of \autoref{eq:weight_func}, but is integrated along the line of sight via \autoref{eq:convergence}.\footnote{In fact, the convergence is the Laplacian of the gravitational potential on the image plane, which accounts for the terms in front of $\tilde n^i(\chi)$ in \autoref{eq:convergence}: the first two are due to the Poisson equation to go from the potential to the density, and the third one is the Laplacian in harmonic space.}
\item Each contribution is projected according to its specific spherical Bessel function---e.g.\ for RSDs it is derived twice, because it is a projected radial derivative.
\end{itemize}

If we are interested in constraining standard cosmological parameters, the lowest multipoles, corresponding to ultra-large scales, are of little interest \citep{Camera:2014sba,Lorenz:2017iez}. This allows us to resort to the Limber approximation, thus getting rid of the integration of the spherical Bessel functions, which is computationally expensive and highly oscillating, thus inducing numerical instabilities. It is worth noting, however, that there are nowadays publicly available routines implementing fast Fourier transforms, such as \angpow\ \citep{Camp}, which can be applied for the computation of tomographic power spectra beyond the Limber approximation in the case one was interested to the largest scales or wanted to reduce the multipole cuts for cross-bin correlations, \citep[see also][]{Chisari_2019}.

The Limber approximation works well for $\ell\gg1$, and the spherical Bessel functions are replaced by a Dirac Delta, viz.\
\begin{equation}
j_\ell(k\chi)\underset{\ell\gg1}{\longrightarrow}\sqrt{\frac{\upi}{2\ell+1}}\delta_{\rm D}\left(\ell+\frac{1}{2}-k\chi\right).
\end{equation}
By implementing this into \autoref{eq:Cl_fullsky}, we get:
\begin{multline}
C^{\rm g,den+mag}_{\ell\gg1}(z_i,z_j)= \\
\int\de\chi\,\frac{W^i_{\rm g}(\chi)W^j_{\rm g}(\chi)}{\chi^2}P_{\rm lin}\left(k=\frac{\ell+1/2}{\chi}\right),
\label{eq:Cldenmag_Limber}
\end{multline}
with
\begin{equation}
W^i_{\rm g}(\chi)=W^i_{\rm g,den}(\chi)+W^i_{\rm g,RSD}(\chi)+W^i_{\rm g,mag}(\chi).\label{eq:W_tot}
\end{equation}
Here, we have split the contributions into three separate window functions: the standard one, for density fluctuations,
\begin{equation}
    W^i_{\rm g,den}(\chi)=n^i(\chi)b(\chi)D(\chi);
    \label{eq:W_den}
\end{equation}
the one for RSDs, found in Paper I to be
%\begin{align}
%    W^i_{\rm g,RSD}(\chi)=\frac{2\ell^2+2\ell-1}{(2\ell-1)(2\ell+3)}\left[n^ifD\right](\chi)  \nonumber \\ %n^i(\chi)f(\chi)D(\chi)\\ 
%    -\frac{(\ell-1)\ell}{(2\ell-1)\sqrt{(2\ell-3)(2\ell+1)}}\left[n^ifD\right]\left(\frac{2\ell-3}{2\ell+1}\chi\right)\nonumber\\
%    -\frac{(\ell+1)(\ell+2)}{(2\ell+3)\sqrt{(2\ell+1)(2\ell+5)}}\left[n^ifD\right]\left(\frac{2\ell+5}{2\ell+1}\chi\right);
%    \label{eq:W_RSD}
%\end{align}

\begin{multline}
    W^i_{\rm g,RSD}(\chi)=\frac{2\ell^2+2\ell-1}{(2\ell-1)(2\ell+3)}\left[n^ifD\right](\chi)  \\ %n^i(\chi)f(\chi)D(\chi)\\ 
    -\frac{(\ell-1)\ell}{(2\ell-1)\sqrt{(2\ell-3)(2\ell+1)}}\left[n^ifD\right]\left(\frac{2\ell-3}{2\ell+1}\chi\right)\\
    -\frac{(\ell+1)(\ell+2)}{(2\ell+3)\sqrt{(2\ell+1)(2\ell+5)}}\left[n^ifD\right]\left(\frac{2\ell+5}{2\ell+1}\chi\right);
    \label{eq:W_RSD}
\end{multline}
and that of magnification bias,\footnote{Note that the multipole factors in \autoref{eq:W_mag} are usually omitted in the literature when describing magnification in the Limber approximation, as easy to see that they cancel each other out in the limit $\ell\gg1$.}
\begin{multline}
    W^i_{\rm g,mag}(\chi)=\\\frac{3\ell(\ell+1)}{(\ell+1/2)^2}
    \om\ho^2\left[1+z(\chi)\right]\chi^2\tilde n^i(\chi)\left[\Q(\chi)-1\right]D(\chi)
%    \int_\chi^\infty\de\chi^\prime\,n^i(\chi^\prime)\frac{\chi^\prime-\chi}{\chi^\prime}.
    \label{eq:W_mag}
\end{multline}

\section{Survey specifications}
\label{sec:EMU_distribution}
As mentioned in \autoref{sec:intro}, we decide to focus on radio continuum surveys, because they are an ideal test case for magnification, thanks to their unrivalled depth. The NRAO VLA Sky Survey \citep[NVSS,][]{Condon_1998} has been the primary source of data for previous cosmological analyses based on radio continuum galaxies \citep[e.g.][]{PhysRevLett.88.021302,refId0,Boughn,Nolta_2004,PhysRevD.76.043510,Raccanelli,PhysRevD.78.043519,PhysRevD.78.123507,10.1111/j.1365-2966.2011.19200.x,Rubart,PhysRevD.89.023511,Nusser_2015,PlanckXXI}. The potentiality of oncoming radio continuum surveys for cosmology has also been extensively studied in recent years \citep{10.1111/j.1365-2966.2012.20634.x,10.1111/j.1365-2966.2012.22073.x,Raccanelli_2015,Bertacca_2011,Jarvis:2015asa,10.1093/mnras/stv040,10.1093/mnras/stu1015,10.1093/mnras/sty1029,Scelfo_2018,Ballardini_2018,Bernal_2019}.

For the present analysis, we adopt the specifications of the Evolutionary Map of the Universe (EMU). EMU is a deep radio-continuum full-sky survey \citep{norris2011} at ASKAP \citep{johnston2007,Johnston2008}, whose goal is to detect extragalactic objects in the continuum across the  entire southern sky, up to $\delta=+30^\circ$. Even though ASKAP was designed as a precursor to the SKA, the large field of view, accurate pointing and angular resolution, and sensitive phased-array feeds will render it the foremost radio survey instrument in the frequency range around 1 GHz during the next decade. The EMU survey, covering such a wide area, and going much deeper than previous large-area radio continuum surveys, will be able to map the large-scale distribution of matter over a larger volume than has previously been possible, and so will be ideal to investigate extensions of the \lcdm\ model \citep{10.1111/j.1365-2966.2012.20634.x,10.1111/j.1365-2966.2012.22073.x,Bernal_2019}. 

EMU will cover an area of $30,000\,\mathrm{deg}^2$ with a sensitivity of $10\,\mu {\rm Jy}$ per beam rms, and a resolution of $\sim10\,\mathrm{arcsec}$, over the frequency range of 800-1400 MHz. To estimate the redshift distribution $n(z)$ of active galactic nuclei and star-forming galaxies, a 10-sigma detection limit of 100 $\mu {\rm Jy}$ is assumed, and galaxies are sampled from the mock catalogues generated by the SKA Simulated Skies (S-cubed)\footnote{http://s-cubed.physics.ox.ac.uk/} simulations down to that limit. The distribution of redshifts and magnitudes from these mocks are used to estimate the overall $n(z)$, and also the magnification bias, $\Q(z)$.

Under the assumption that additional external data will be available for the redshifts of part of EMU galaxies (e.g.\ cross-identifications, \citealt{McAlpine:2012cu}; Bayesian hierarchical models, \citealt{Harrison:2017pcu}; or so-called clustering redshifts, \citealt{Menard:2013aaa}), we here scrutinise two binning scenarios. The former, in which we assume we can differentiate only between low- and high-redshift galaxies (divide set at $z=1$), is more conservative; the latter sees five redshift bins, four of which of width $\Delta z=1$ below $z=2$, and the fifth collecting all the galaxies above. The expected numbers for these settings are given in \autoref{tab:2bins} and \autoref{tab:5bins}.
\begin{table}
\centering
\caption{Estimated number densities, galaxy bias, and magnification bias for EMU sources grouped in 2 redshift bins.}
 \begin{tabular}{c c c c c c}
 \hline
 Bin &  $z_{\rm min}$ & $z_{\rm max}$ & \# of gal. ($\times10^6$)  & bias & mag. bias \\  
 \hline
 1 & 0.0 & 1.0 & 10.68 & 0.833 & 1.050 \\ 
 2 & 1.0 & 6.0 & 11.58 & 2.270 & 1.298 \\
 \hline
\end{tabular}
\label{tab:2bins}
\end{table}
\begin{table}
\centering
\caption{Same as \autoref{tab:2bins}, but for EMU sources grouped in 5 redshift bins.}
 \begin{tabular}{c c c c c c} 
 \hline
 Bin &  $z_{\rm min}$ & $z_{\rm max}$ & \# of gal. ($\times10^6$)  & bias & mag. bias \\  
\hline
 1 & 0.0 & 0.5 & 5.55 & 1.000 & 0.953 \\ 
 2 & 0.5 & 1.0 & 5.13 & 1.124 & 1.273 \\ 
 3 & 1.0 & 1.5 & 4.43 & 1.920 & 1.569 \\
 4 & 1.5 & 2.0 & 2.70 & 3.250 & 1.176 \\
 5 & 2.0 & 6.0 & 4.05 & 4.046 & 0.964 \\
 \hline
\end{tabular}
\label{tab:5bins}
\end{table}

We discussed above that radio continuum surveys lack information in redshift and therefore the most realistic representation of the galaxy sampling in redshift space is that of residing in Gaussian bins. However, we decide to consider the case of sharp top-hat bins which are not correlated in redshift. We apply this mostly for the sake of fully exploring the potential of magnification. The magnification bias is expected to induce a correlation even between uncorrelated redshift bins, in a sense that the lower-$z$ bins are the `lenses' and the high-$z$ bins the `sources'. Thus, it is worth investigating magnification in this case, too, implemented at least at the \lcdm\ scenario.

\begin{figure*}
\centering
\includegraphics[width=0.45\textwidth,trim={2cm 9cm 4cm 7cm},clip]{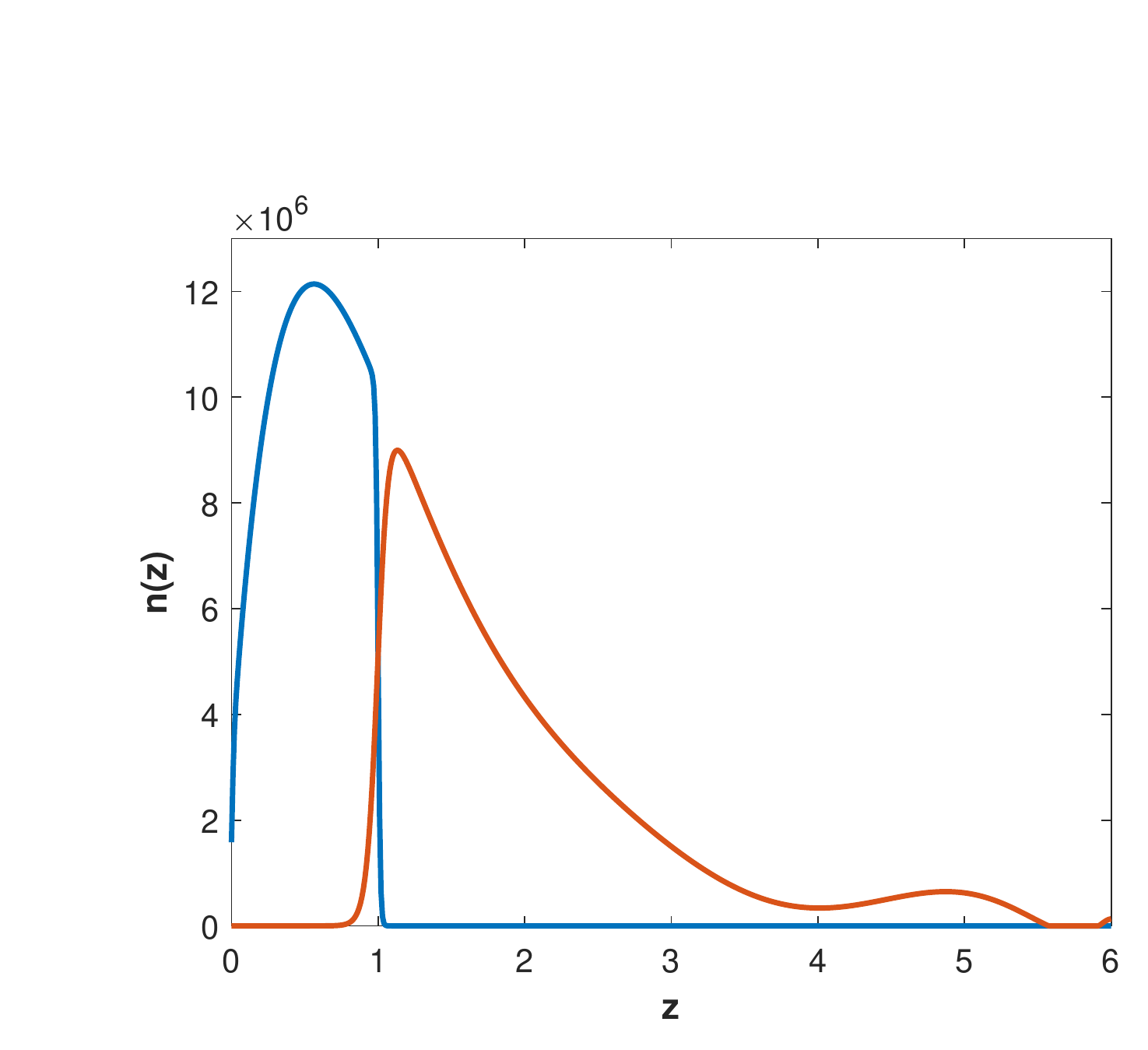}\includegraphics[width=0.45\textwidth,trim={2cm 9cm 4cm 7cm},clip]{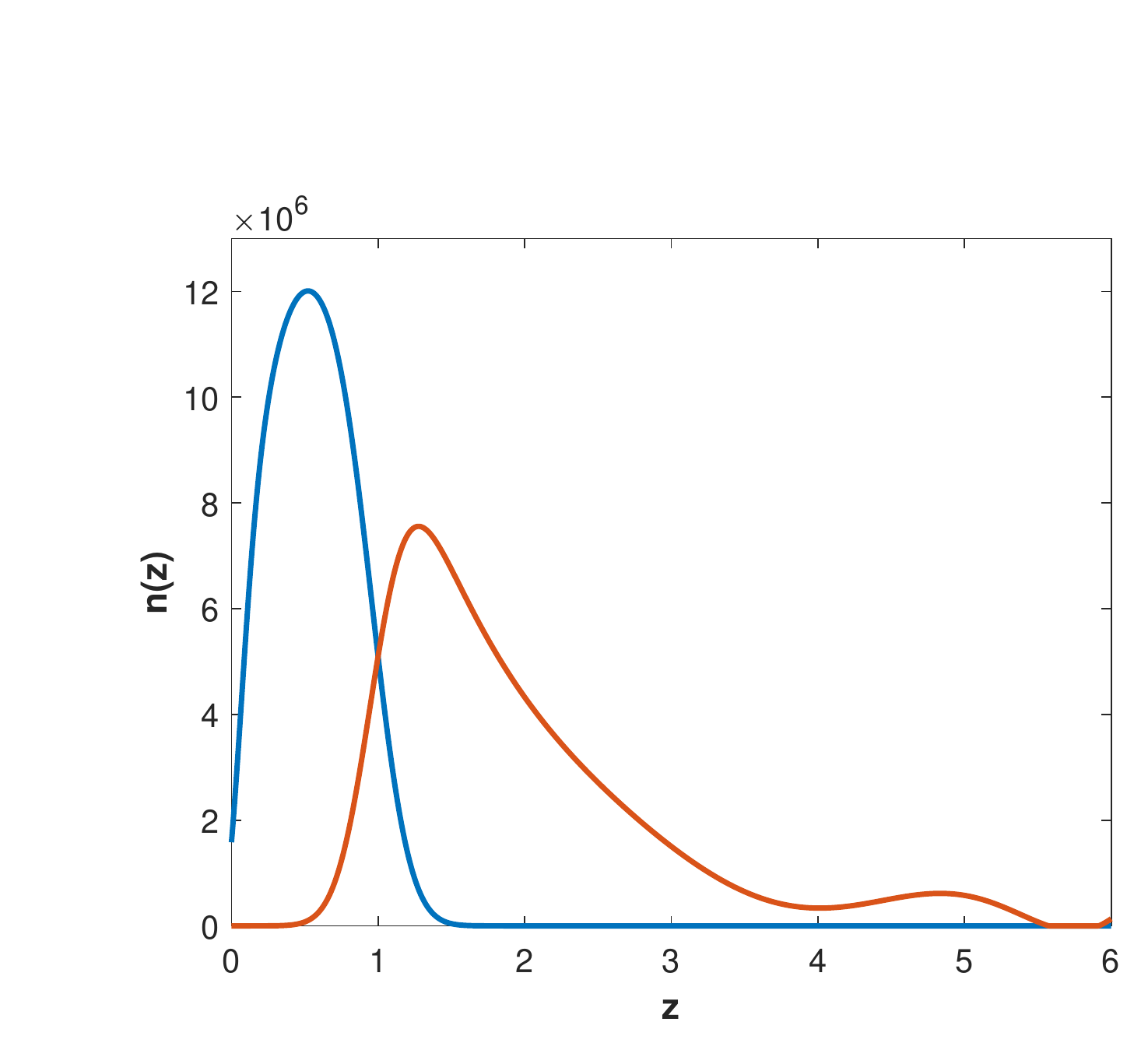}\\\includegraphics[width=0.45\textwidth,trim={2cm 9cm 4cm 9.5cm},clip]{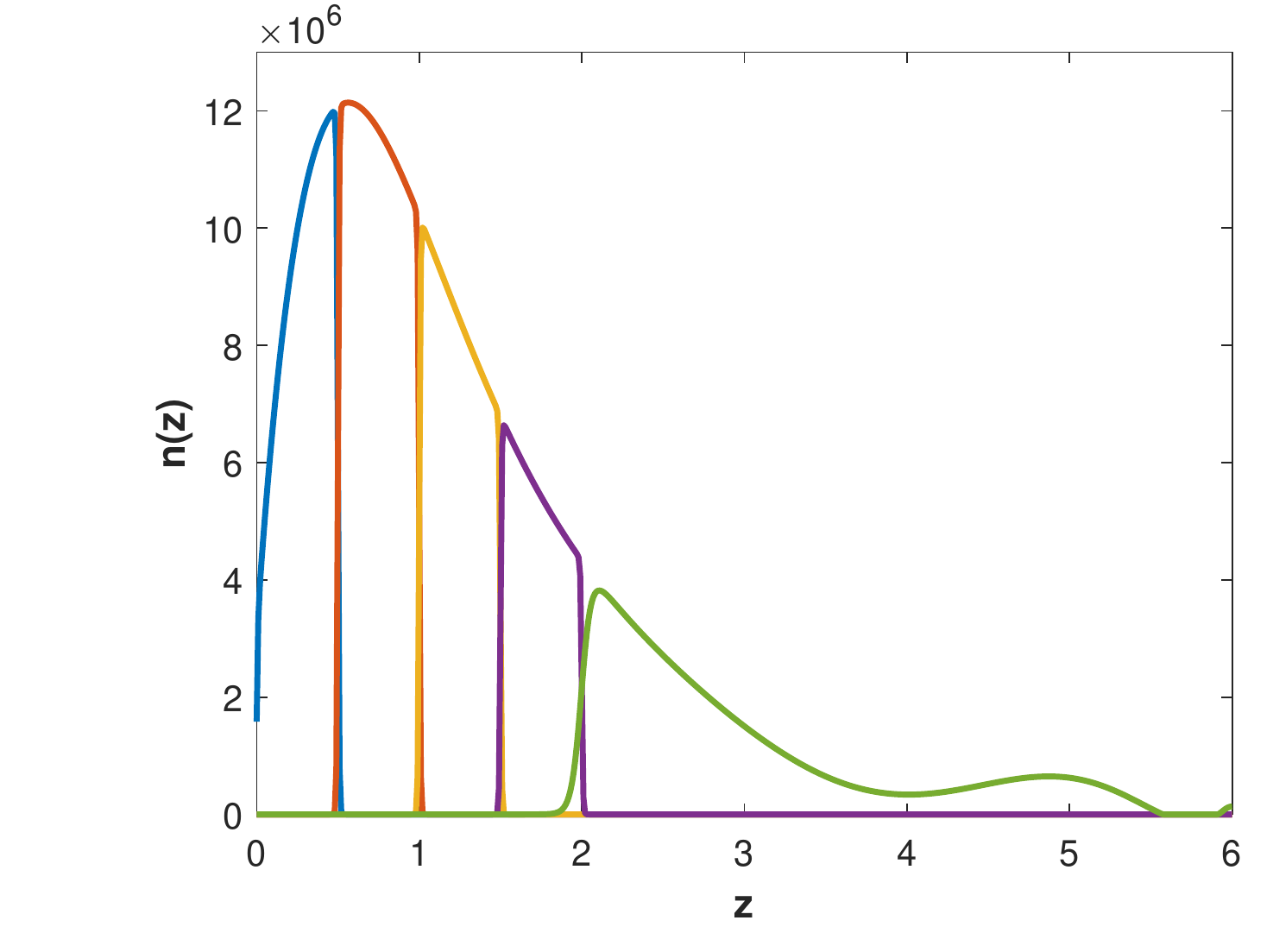}\includegraphics[width=0.45\textwidth,trim={2cm 9cm 4cm 9.5cm},clip]{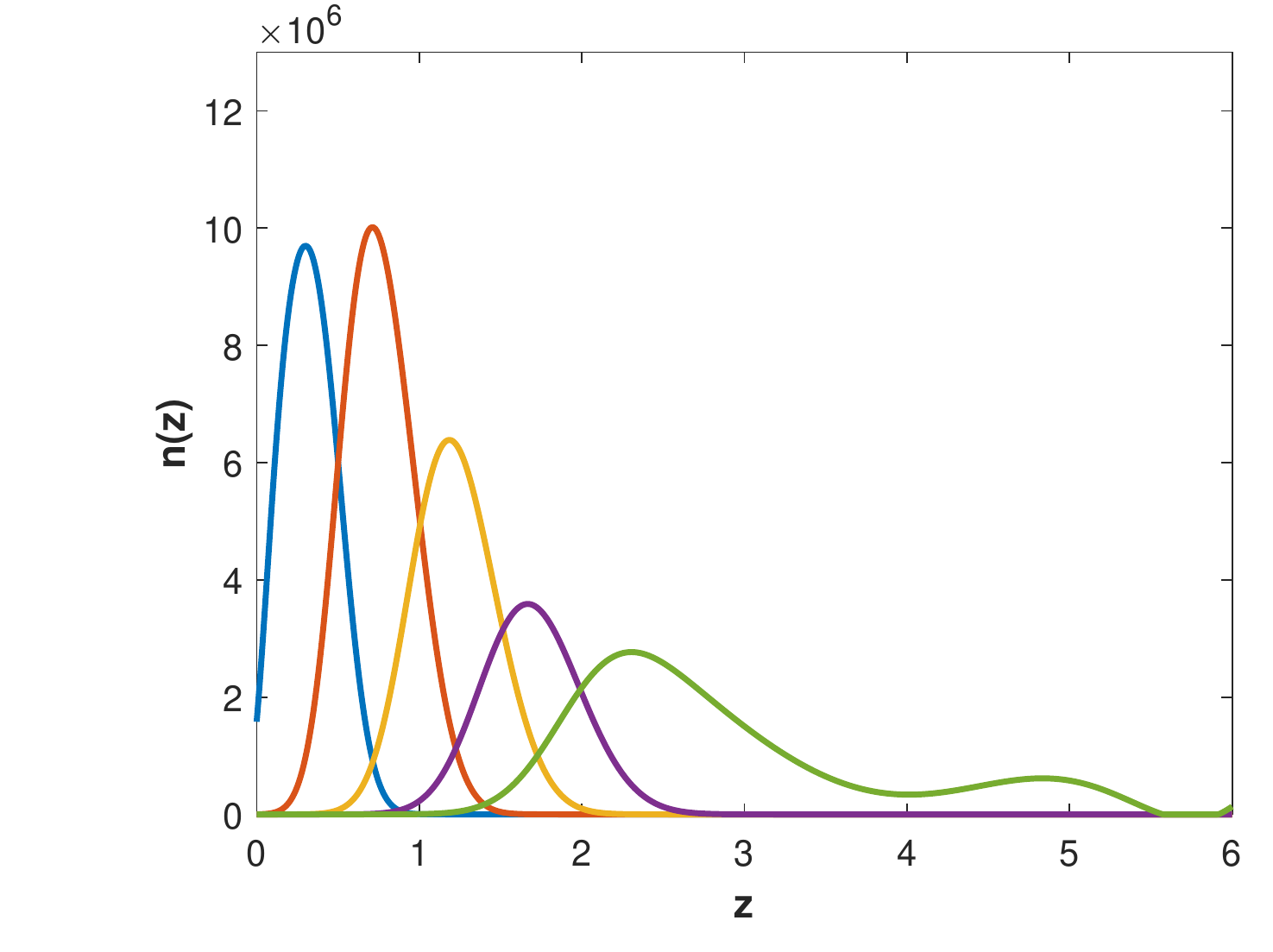}
\caption{The EMU galaxy redshift distribution for top-hat (left panels) and Gaussian (right panels) binning. The top and bottom panels present the 2 and the 5 bins, respectively.}
\label{fig:distribution}
\end{figure*}

Given that $\de N$ is the number of galaxies inside a bin of width $\de z$, the redshift distribution of sources is $N(z)=\de N/\de z$.\footnote{The number of sources has also been calculated in very narrow 32 redshift bins, which are not shown here for clarity.} Then, the $N(z)$ points are fitted with a 7th order polynomial, $n(z)$, by which we denote the total number counts of sources with redshift. The distribution of sources residing in the $i$th bin thus is $n^i(z)$, and the angular number counts of galaxies reads
\begin{equation}
    \bar n^i=\int\de z\,n^i(z),
    \label{eq:n_i}
\end{equation}
Therefore, the total number counts of galaxies is simply $\bar n=\sum_i{\bar n^i}$.\footnote{Note again that $n^i(z)$ is normalised to unity in the equations of the previous section, like \autoref{eq:weight_func}, \autoref{eq:W_den}, \autoref{eq:W_RSD}, and \autoref{eq:W_mag}, meaning that it as to be read as $n^i(z)/\bar n^i$.} The final, fitted redshift distributions, convolved with the bins, are shown in \autoref{fig:distribution}.

Top-hat bins (left panels of \autoref{fig:distribution}) have been modelled as
\begin{equation}
    n^i(z)=\frac{1}{2}\left\{1-\tanh\left[\frac{|z-\bar z_i|-\sigma}{r\sigma}\right]\right\},
\end{equation}
where $\bar z_i$ is the centre of the $i$th bin, $\sigma$ the half top-hat width, and $r$ the smoothing edge, which we fix to 0.03. The smearing ensures the numerical stability in the integration over the bin. Instead, to model Gaussian bins, we consider the ranges $z_{\rm min}$ and $z_{\rm max}$ of \autoref{tab:2bins} and \autoref{tab:5bins}, and definde
\begin{multline}
    n^i(z)=\\
    \frac12 n(z) \left[\operatorname{erfc}\left(\frac{z^i_{\rm min}-z}{\sqrt2 \sigma(z^i_{\rm min})}\right)-\operatorname{erfc}\left(\frac{z^i_{\rm max}-z}{\sqrt2 \sigma(z^i_{\rm max})}\right) \right].
\end{multline}
Note that, in this latter case, we introduce a redshift dependence of the scatter of the distribution, $\sigma(z)$. Specifically, we adopt a quite large uncertainty, $\sigma(z)=0.1(1+z)$. The Gaussian bins are shown in the right panels of \autoref{fig:distribution}.

\section{Cosmological models}
\label{sec:cosmo_theory}
In this work, we will investigate the vanilla \lcdm\ model and two of its most popular extensions: the case of a dynamical dark energy component and a phenomenological modified gravity model. All the model parameters are summarised in \autoref{tab:params}. For \lcdm, 
%
%\subsection{$\Lambda$CDM}
%We introduce a baseline fiducial cosmology with values as seen in \autoref{tab:params}. For this model, 
we present the constraints for the parameter set $\{\om,\,h,\,\sigma_8\}$ alone, whilst the other parameters are fixed to their fiducial values.   

\subsection{Dark energy}
\label{sec:DE}
The first extension to the \lcdm\ model is a dynamical dark energy model (DE, hereafter), where the dark energy equation of state is not constant throughout the cosmic history, but it is rather allowed to evolve with time. According to \citep{doi:10.1142/S0218271801000822,PhysRevLett.90.091301}, by Taylor expanding an evolving dark energy equation of state and keeping only the first order term we have
\begin{equation}
    w_{\rm DE}(z)=w_0+w_a\frac{z}{1+z}.
\end{equation}
Therefore, we add to the \lcdm\ model the parameter set both $w_0$ and $w_a$.

\subsection{Modified gravity}
\label{sec:MG}
An alternative explanation for the late-time accelerated cosmic expansion is offered by modified gravity theories (MG, hereafter). This approach sees the effects we ascribe to dark energy (and even dark matter) are in fact due to our wrong interpretation of the data in a regime where general relativity no longer holds \citep{Clifton:2011jh}. For the purpose of our paper, we assume a popular phenomenological parameterisation accounting for the peculiar effect of modified gravity on structure formation \citep{Amendola:2007rr,Zhao:2010dz,PhysRevD.92.023003}.
%Starting from the general relativity, one can derive the perturbed FRWL metric in the conformal Newtonian gauge:
%\begin{equation}
%    ds^2 = a(\eta)^2 [-(1+2\Psi)d\eta^2+(1-2\Phi)d\chi^i d\chi_i],
%\end{equation}
%where $\eta$ is the conformal time, $\chi_i$ the comoving coordinates, and $\Psi$,  $\Phi$ the scalar potentials. The matter can feel both the Newtonian potential $\Psi$ and the lensing potential $\Phi$, while the relativistic particles can only feel the latter. The potential can be modified by introducing a term $Q_0$ in the Poisson equation: 
Specifically, we can assume a modified Poisson equation
\begin{equation}
    \nabla^2\Phi=4\pi GQ  a^2 \bar\rho\delta ,
\end{equation}\label{eq:MG1}
where $Q$ is in principle a function of space and time, and acts as an effective gravitational constant. Moreover, the two metric potentials can be different, and the function $R$ describes the ratio of the two, viz.\
\begin{equation}
    R = \frac{\Psi}{\Phi}.
\end{equation}\label{eq:MG2}
Thus, we add as free parameters the two present-values of these quantities, $Q_0$ and $R_0$. In fact, given that they are degenerate, it is very convenient to define the derived parameter $\Sigma_0=Q_0(1+R_0)/2$, and therefore use the parameter set $\{Q_0,\,\Sigma_0\}$ instead, along with the parameters of the \lcdm\ model. 

\begin{table*}
\centering
\caption{Prior ranges and fiducial values on the nuisance and cosmological parameters (\lcdm\ best-fit of \citealt{Ade2015}). Some parameters are purposely allowed to have wider or narrower prior ranges due to the difference in the constraining power of the results depending on the number of the bins considered. (When two sets of values are present, values in parentheses refer to the 5 bin case, as opposite to those outside that are relative to the 2 bin case.)}
\begin{tabularx}{\textwidth}{Xllll}
	\hline
	 Parameter description & Parameter symbol & Fiducial value & Prior type & Prior range \\
	\hline
	\hline
	Present-day fractional matter density & $\om$ & 0.3089 & Flat & $[0.1,0.6]$\\
	Dimensionless Hubble parameter & $h$ & 0.6774 & Flat for 2(5) bins & $[0.3,1.0]$$([0.5, 1.0])$\\
    Amplitude of clustering$^\ddag$ & $\sigma_8$ & 0.8159 & Flat for 2(5) bins & $[0.4,1.4]$$([0.5,1.2])$\\
	\hline
    Present-day fractional baryon density & $\ob$ & 0.0486 & -- & --\\
    Slope of the primordial curvature power spectrum & $\ns$ & 0.9667 & -- & --\\
	Amplitude of the primordial curvature power spectrum$^\ddag$ & $\ln(10^{10}\As)$ & $3.064$ & -- & --\\
    Optical depth to reionisation & $\tau_{\rm re}$ & 0.066 & -- & --\\
	\hline
    Bias amplitude parameter for the whole redshift range$^\P$ & $\alpha_{\rm EMU}$ & 1.0 & Flat & $[0.4,1.6]$ \\
	\hline
	 Free bias amplitude in each redshift bin$^\S$ & $b_i$ $i=1\ldots2(5)$ & See \autoref{tab:2bins}(\autoref{tab:5bins}) & Flat for 2(5) bins & $[0.1,3.5]$$([0.1,9.0])$ \\
	 \hline 
	 Present-day dark energy equation of state & $w_0$ & $-1.0$ & Flat & $[-3.0,2.0]$ \\
	 Dark energy evolution parameter & $w_a$ & 0.0 & Flat & $[-6.0,4.0]$ \\
	 \hline	
	 Modified gravity parameter & $Q_0$ & 1.0  & Flat & $[0.0,8.0]$ \\
	 Modified gravity parameter & $R_0$ & 1.0 & Flat & $[-1.0,8.0]$ \\
	 \hline
\end{tabularx}\label{tab:params}
\raggedright\footnotesize{$^\ddag$ Instead of setting the prior on the parameter $\As$ accounting for the matter perturbations amplitude, we opt for $\sigma_8$, following the convention in LSS.\\
$^\P$ The prior range reported on the parameter is applied in the `realistic' scenario alone (notation mirrors Paper I).\\
$^\S$ The prior range reported on the parameter is applied in the `conservative' scenario alone.}
\end{table*}

\section{Methodology}
\label{sec:like}
To forecast constraints on cosmological parameters, we follow a likelihood-based approach. The first step is to estimate the covariance matrix, $\bm\Gamma_{\ell\ell^\prime}$, for our observable, namely the galaxy clustering power spectrum in harmonic space given in \autoref{eq:Cldenmag_Limber}. We use the analytical form of the Gaussian covariance matrix, as already implemented in \cosmosis, with the following entries
\begin{multline}
\Gamma^{ij,kl}_{\ell\ell^\prime}=\\
\frac{\delta_{\rm K}^{\ell\ell^\prime}}{2\ell\Delta\ell f_{\rm sky}}[\widetilde C^{\rm g}_\ell(z_i,z_k)\widetilde C^{\rm g}_\ell(z_j,z_l)
+\widetilde C^{\rm g}_\ell(z_i,z_l)\widetilde C^{\rm g}_\ell(z_j,z_k)],\label{eq:covmat}
\end{multline}
where $f_{\rm sky}$ the fraction of the sky covered by the survey, $\Delta \ell$ the multipole range, $\delta_{\rm K}$ the Kronecker symbol, and
\begin{equation}
\widetilde C^{\rm g}_\ell = C^{\rm g}_\ell + \frac{\delta_{\rm K}^{ij}}{\bar n^i},
    \label{eq:noise}
\end{equation}
is the observed signal---namely, signal plus shot noise, with $\bar n^i$ defined in \autoref{eq:n_i}.\footnote{In the denominator of \autoref{eq:covmat} we use the notation of \citet{JoachimiBridle2010} and keep $2\ell$ instead of $(2\ell+1)$. This makes no difference for our results, since we are at the Limber limit allowing scales $\ell\gg1$. Also, the analysis is based on the effect of neglecting the magnification bias and therefore such a choice can be accepted safely at the cost of no loss of generality.} We employ $N_\ell=20$ multipole bins (see \autoref{sec:cuts} for the range adopted), and for all redshift and multipole bin values we construct the full data vector ${\bm d}_\ell=[{\bm C}^{\rm g}_\ell]$, as well as the theory vector ${\bm t}_\ell({\bm \theta})$, which is a function of the parameter set, $\bm\theta$. With all the above one can construct the $\chi^2$ as
\begin{equation}
    \chi^2 =\sum_{\ell,\ell^\prime=\ell_{\rm min}}^{\ell_{\rm max}} [{\bm d}_\ell - {\bm t}_\ell({\bm \theta})]^{\sf T} (\bm\Gamma_{\ell\ell^\prime})^{-1} [{\bm d}_\ell - {\bm t}_\ell({\bm \theta})],
    \label{eq:chi2}
\end{equation}
which is to be minimised for some specific values of the parameters. %We should note that there is no need for a normalisation factor in \autoref{eq:chi2}, since the covariance matrix is assumed not to be depend on the parameter set $\theta$. 
Matrix transposition and inversion are denoted by `${\sf T}$' and `$-1$', respectively. 

\subsection{Multipole cuts}
\label{sec:cuts}
Since Limber approximation is valid only at $\ell \gg 1$, we have derived the \lm\ below which we can trust no longer the angular power spectra values computed via \autoref{eq:Cldenmag_Limber}. To do so, we compare results computed by our modified \cosmosis\ code with the full solution of the \class\ Boltzmann solver and keep only the multipoles where the relative error between the two codes is below 5\% (see Paper I). We make this chose since this percentage offset is within the standard deviation of the signal measurement. 

Additionally, we apply an upper cut at $\ell_{\rm max}= \chi(\bar z_i) k_{\rm max}$, since we ignore the nonlinear scales in our analysis. Here, $\bar z_i$ is the centre of the $i$th redshift bin, whilst the maximum wavenumber is chosen to be $k_{\rm max}=\pi / (2R_{\rm min})$, where $R_{\rm min}$ is the radius of a sphere inside which the over-density fluctuations at $z=0$ have a value
\begin{equation}
\sigma^2(R)=\frac{1}{2\pi^2}\int{\de k}\,k^2P_{\rm lin}(k)\left|W(k R)\right|^2,
\label{eq:highmul}
\end{equation}
with the spherical top-hat function being $W(x)=3j_1(x)/x$. The matter density variance is chosen to be $\sigma^2(R_{\rm min})=1$, yielding $k_{\rm max}=0.25\,\,h\,{\rm Mpc}^{-1}$.

The \lm\ and \lM\ cuts are applied to each bin pair according to the all the configurations of the EMU distribution (see again \autoref{fig:distribution}), and are shown in  \autoref{tab:multipoles}, where RSDs do not appear explicitly because we found that their inclusion does not affect the value of \lm. (On the other hand, \lM\ does not depend on the terms included in \autoref{eq:Cldenmag_Limber}, as it is only a function of $k_{\rm max}$ and the central redshift of the bin.)
\begin{table*}
\centering
\caption{The $\ell_{\rm min}$ and $\ell_{\rm max}$ values for all the EMU bin configurations. The former is specified as the point where the relative error between \cosmosis\ and \class\ angular power spectra measurements is below 5\%, while the latter in the limit where $\ell_{\rm max}=k_{\rm max} \chi(\bar z_i)$ with $\bar z_i$ the centre of the $i$th bin.}
\begin{tabular}{ccccccccccccc}
\hline
\multicolumn{6}{c}{2 redshift bins} && \multicolumn{6}{c}{5 redshift bins} \\
\cline{1-6}\cline{8-13}
\multicolumn{5}{c}{\lm} & \multicolumn{1}{c}{\lM} && \multicolumn{5}{c}{\lm} & \multicolumn{1}{c}{\lM} \\
\cline{1-5}\cline{8-12}
\multicolumn{2}{c}{Top-hat} && \multicolumn{2}{c}{Gaussian} &&& \multicolumn{2}{c}{Top-hat} && \multicolumn{2}{c}{Gaussian} & \\
w/o mag & w/ mag && w/o mag & w/ mag &&& w/o mag & w/ mag && w/o mag & w/ mag & \\
\hline
\hline
3 & 2 && 2 & 2 & 480 && 2 & 2 && 2 & 2 & 257 \\

10 & 12 && 10 & 10 & 1718 && 6 & 6 && 8 & 8 & 673 \\

$-$ & $-$ && $-$ & $-$ & $-$ && 17 & 18 && 11 & 11 & 982 \\

$-$ & $-$ && $-$ & $-$ & $-$ && 24 & 25 && 10 & 10 & 1215 \\

$-$ & $-$ && $-$ & $-$ & $-$ && 24 & 25 && 9 & 9 & 1813 \\
\hline
\end{tabular}
%\begin{tabular}{ccccccccccccc}%{lllllllllllll}
%\hline
%\multicolumn{3}{c}{2 top-hat bins} & \multicolumn{3}{c}{5 top-hat bins} && \multicolumn{3}{c}{2 Gaussian bins} & \multicolumn{3}{c}{5 Gaussian bins} \\
%\cline{1-6}\cline{8-13}
%\multicolumn{2}{c}{\lm} & \multicolumn{1}{c}{\lM} & \multicolumn{2}{c}{\lm} & \multicolumn{1}{c}{\lM} && \multicolumn{2}{c}{\lm} & \multicolumn{1}{c}{\lM} & \multicolumn{2}{c}{\lm} & \multicolumn{1}{c}{\lM} \\
%\cline{1-2}\cline{4-5}\cline{8-9}\cline{11-12}
%w/o mag & w/ mag & & w/o mag & w/ mag & && w/o mag & w/ mag & & w/o mag & w/ mag & \\
%%den/den+RSD & den+mag & & den/den+RSD & den+mag & && den/den+RSD & den+mag & & den/den+RSD & den+mag & \\
%\hline
%\hline
%3 & 2 & 480 & 2 & 2 & 257 && 2 & 2 & 480 & 2 & 2 & 257 \\
%
%10 & 12 & 1718 & 10 & 10 & 673 && 6 & 6 & 1718 & 8 & 8 & 673 \\
%
%$-$ & $-$ & $-$ & 17 & 18 & 982 && $-$ & $-$ & $-$ & 11 & 11 & 982 \\
%
%$-$ & $-$ & $-$ & 24 & 25 & 1215 && $-$ & $-$ & $-$ & 10 & 10 & 1215 \\
%
%$-$ & $-$ & $-$ & 24 & 25 & 1813 && $-$ & $-$ & $-$ & 9 & 9 & 1813 \\
%\hline
%\end{tabular}
\label{tab:multipoles}
\end{table*}

\section{Results and discussion}
\label{sec:results}
Let us summarise again here the cosmological parameter sets for the three different cosmological models, $\bm \theta_\textrm{\lcdm}=\{\om,\,h,\,\sigma_8\}$, ${\bm \theta_{\rm DE}}=\bm \theta_\textrm{\lcdm}\cup\{w_0,\,w_a \}$ and ${\bm\theta_{\rm MG}}=\bm \theta_\textrm{\lcdm}\cup\{Q_0,\,\Sigma_0 \}$. In our forecasting analysis, we use the Bayesian sampler \Multinest\ \citep{FHB2009}. 

We forecast cosmological parameter constraints using both the incomplete $C^{\rm g,den}_{\ell\gg1}$ and the correct $C^{\rm g,den+mag}_{\ell\gg1}$ spectra for the different binning configurations of EMU, fitting the mock data using a likelihood of the form described in \autoref{sec:like}. Note that for the moment we neglect RSDs in the modelling of the synthetic data. The reason for this will be come clear afterwards, and we discuss the issue in \autoref{sec:RSD}. The mock-data vector ${\bm d}_\ell$ is thus constructed assuming the density perturbations and the magnification bias described in \autoref{sec:Apower spectrum}, according to the fiducial cosmology given in \autoref{tab:params}.

Additionally, we need to add a number of extra nuisance parameters to our analysis, that will be marginalised over, in addition to the cosmological parameters of interest. These nuisance parameters model our ignorance on some underlying quantity such as the galaxy bias, and depend also upon the binning strategy adopted. We introduce three cases:
\begin{enumerate}
    \item An idealistic scenario, where the galaxy bias is perfectly known, keeping its fiducial values as in \autoref{tab:2bins} and \autoref{tab:5bins};
    \item A realistic scenario, with an single bias amplitude parameter spanning the whole redshift range, which is taken as a free parameter;
    \item A realistic, yet conservative scenario, allowing for a free galaxy bias parameter per each redshift bin.
\end{enumerate}

Let us finally remark that the magnification bias for each redshift bin keeps its fiducial value as in \autoref{tab:2bins} and \autoref{tab:5bins}, and it remains fixed throughout the analysis and for all the scenarios. Moreover, we choose to take the means of the posterior distribution instead of the best-fit values to allow for safer conclusions in the case of highly non-Gaussian posterior distributions (see Paper I). The results are presented and discussed thoroughly in the next subsections where we uniformly opt to show the constraints on the derived parameter
\begin{equation}
    S_8=\sigma_8 \sqrt{\frac{\om}{0.3}},
\end{equation}
which is better constrained than $\sigma_8$, and is not correlated with $\om$. In all plots the means of the posterior along with the 68\% marginal errors for each parameter are shown.

\subsection{Constraints on $\mathbf\Lambda$CDM}
\label{sec:TH_Lamblda}
\begin{figure*}
\centering
\includegraphics[width=0.45\textwidth,trim={3.6cm 9cm 4cm 9.5cm},clip]{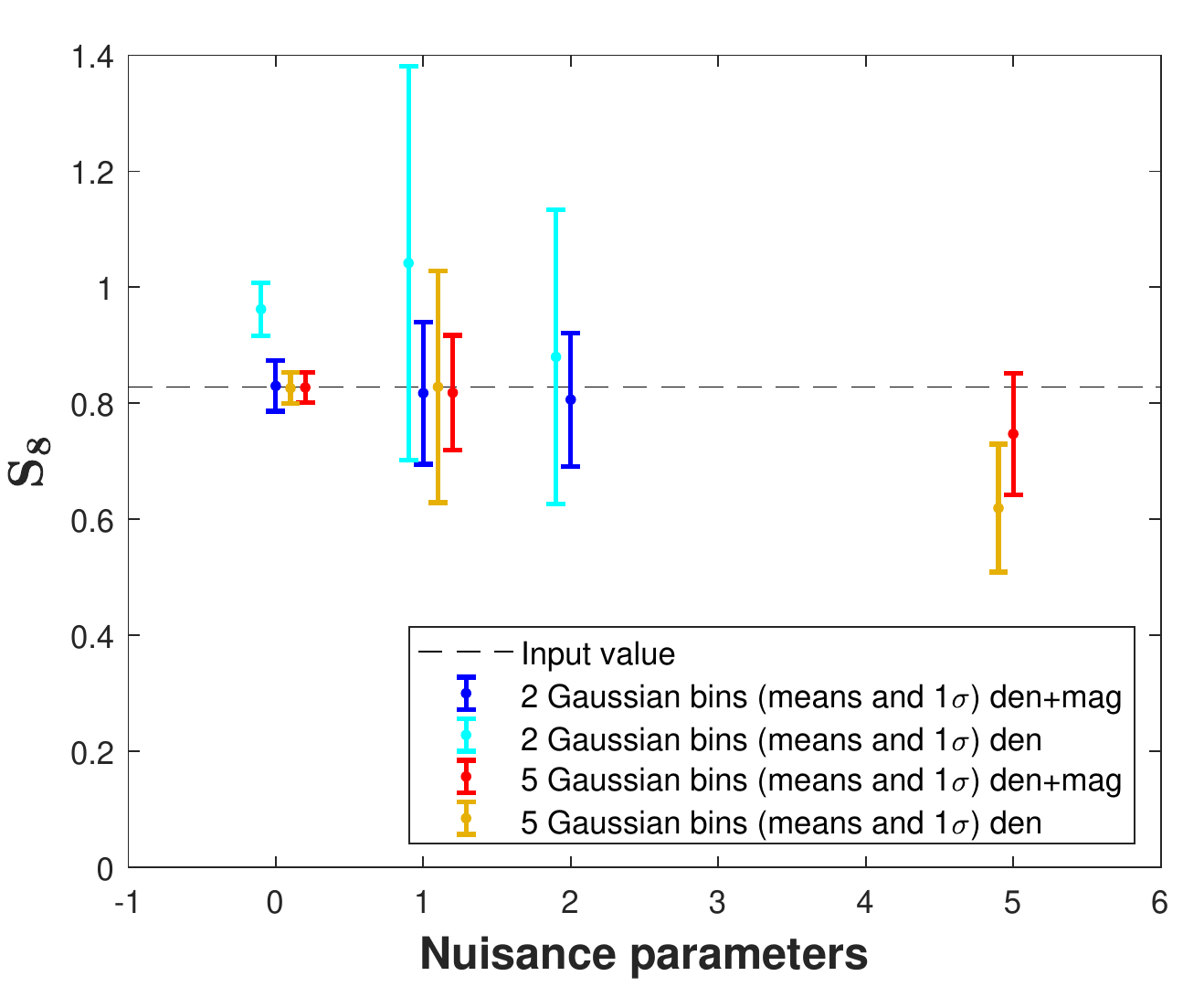}\includegraphics[width=0.45\textwidth,trim={3.6cm 9cm 4cm 9.5cm},clip]{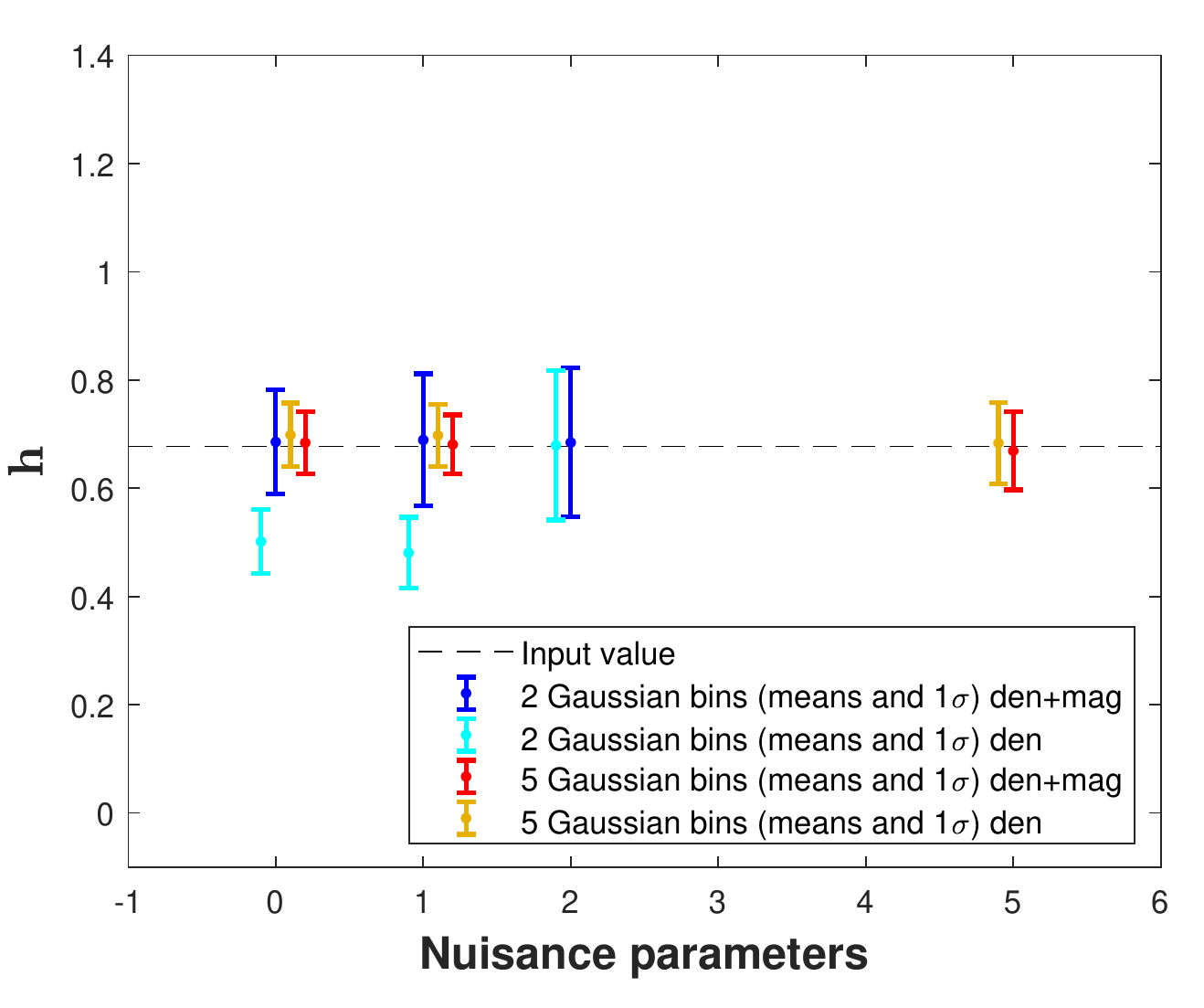}
\caption{EMU mean and 68\% confidence intervals on the derived $S_8$ (left) and $h$ (right) cosmological parameter for Gaussian binning as a function of the number of nuisance parameters for the \lcdm\ model. Note the different colours accounting for the number of bins and the combination of density and magnification in the theory vector.}
\label{fig:LCDM}
\end{figure*}

In \autoref{fig:LCDM} we present the $68\%$ marginal confidence intervals and the means on $\{S_8,\,h\}$ for 2 and 5 Gaussian bins---a binning scenario closer to reality. As a general remark, we shall see that whether we consider the realistic or the conservative scenarios, the constraining power that we get from the correct model (i.e.\ den+mag) is comparable. This is true for both binning configurations, and as we will see in the following sections, this feature remains the same in the cases of extensions of the \lcdm\ model.

Results for top-hat bins are very similar to those obtained with more realistic Gaussian bins, so we report the corresponding figure and tables in \autoref{Top-hat bins}, limiting ourselves to point out that the main difference between Gaussian and top-hat binning is that the latter sees mildly biased estimates for $h$ even in the 5-bin, 1-nuisance parameter case. This is mainly due to the slightly tighter constraints obtained with top-hat bins in this configuration, meaning that the observable is more sensitive to the Hubble constant thanks to the better redshift resolution. Besides this, on a general ground, we see no further, major difference between top-hat and Gaussian bins. This has to be attributed to the fact that the bins considered for the EMU distribution are quite wide regardless of the top-hat or Gaussian bin choice.

Nonetheless, the offsets in the parameter estimates obtained with Gaussian bins are always a bit more pronounced compared to top-hat bins. That is, Gaussian bins, given the poor redshift estimate, are wider than the sharp top-hats, and so have more sources with a significantly different dynamical time, along the line of sight in the same redshift bin. As a result, the wider the bin is, the larger the magnification bias is, inducing a larger offset in the results when excluded.

\subsection{Two Gaussian bins}
In the case where the galaxy bias is perfectly known---the idealistic scenario, marked by `0 nuisance parameters' on abscissas of \autoref{fig:LCDM}---it is evident that when we fit the mock data with the complete model (blue error bar), the input reference values are well within the $68\%$ error interval calculated on both parameter, $S_8$ and $h$. On the other hand, when we assume the incomplete model (cyan error bar), namely ignoring the magnification contribution in the theory vector, the estimates of $\{S_8,\,h\}$ are clearly biased with respect to the input reference.  

Then, in the realistic scenario we introduce a free galaxy bias parameter $\alpha_{\rm EMU}$ for the whole redshift range (`1 nuisance parameters' mark on the $x$-axis).  The results presented on the cosmological set $\{S_8,\,h\}$ are then obtained after marginalising over this nuisance parameter. Interestingly, now the results on $S_8$ are different. That is, even with the incorrect model $S_8$ becomes totally unconstrained (cyan error bar). The reason for this is that the galaxy density field is highly sensitive to the galaxy bias. As a result, there is a degeneracy between the galaxy bias and the amplitude of matter fluctuations, $S_8$. Nonetheless, when we consider magnification, too (blue error bar), we lift this degeneracy considerably, and the error bar shrinks.

Now, we examine the conservative scenario, where we allow for a nuisance bias parameter for each redshift bin, $b_i$, in the range $[0.1,3.5]$ to be marginalised over (`2 nuisance parameters' tick). Constraints on $S_8$ is quite similar to those of the realistic scenario, with the incomplete model yielding a degenerate $S_8$ (cyan error bar) estimate, in turn mitigated by the incorporation of the magnification bias (blue error bar) for the same reasons mentioned above. On the contrary, we see no deviance in the $h$ for the wrong model (cyan error bar). This is probably due to the fact that we use a larger number of nuisance parameters, leading to an overall broadening of the confidence intervals.

The findings for the case of 2 Gaussian bins are quantitatively summarised in \autoref{tab:results_EMU2G_LCDM}.
\begin{table*}
\centering
\caption{Means and corresponding $68\%$ marginal error intervals on cosmological parameters for the EMU radio continuum galaxy survey applying 2 Gaussian bins with the \lcdm\ model.}
\begin{tabular}{lllcllcll}
    \hline
    \multicolumn{9}{c}{2 Gaussian bins ($\Lambda$CDM)} \\
    \hline
   & \multicolumn{2}{c}{Ideal scenario} && \multicolumn{2}{c}{Realistic scenario} && \multicolumn{2}{c}{Conservative scenario} \\
    \cline{2-3}\cline{5-6}\cline{8-9}
   & \multicolumn{1}{c}{den} & \multicolumn{1}{c}{den+mag} && \multicolumn{1}{c}{den} & \multicolumn{1}{c}{den+mag} && \multicolumn{1}{c}{den} & \multicolumn{1}{c}{den+mag} \\
    \hline
    \hline
    $S_8$ & $0.962\pm0.045$ & $0.830\pm0.044$ && $1.04\pm0.34$ & $0.82\pm0.12$ && $0.88\pm0.25$ & $0.81\pm0.12$ \\
    $h$ & $0.502\pm0.059$ & $0.686\pm0.096$ && $0.481\pm0.066$ & $0.69\pm0.12$ && $0.68\pm0.14$ & $0.68\pm0.14$ \\
    \hline
\end{tabular}
\label{tab:results_EMU2G_LCDM}
\end{table*}

\subsection{Five Gaussian bins}
Let us now turn to the results obtained with 5 bins. Starting from the idealistic case, where the galaxy bias is known exactly, it is clear that there is no bias on any cosmological parameter of interest when using the wrong model (yellow error bar). After marginalising over the normalisation bias parameter for the whole redshift range (realistic scenario), a degeneracy between this $\alpha_{\rm EMU}$ and $S_8$ appears (yellow error bar), in a similar fashion to the 2 bin analysis with density only. In agreement with the previous results, the correction of the magnification effect yields more stringent constraints (red error bar). Also, $h$ estimated with the incomplete model (yellow error bar) stays consistent with the fiducial cosmology for both the realistic and the conservative case.

It is worth noting that the picture changes in the conservative case (now allowing this prior range $[0.1,9.0]$) concerning the estimate on $S_8$ with the wrong model (yellow error bar). In detail, this estimate is biased for more than $68\%$ below the reference value. However, the inclusion of magnification corrects for this bias completely (red error bar). The last result on $S_8$ may seem a bit unexpected, as it is evident from the analysis with the 2 bins that both the realistic and the conservative scenarios yield comparable results on $S_8$ that are quite degenerate, yet not biased, with the density-only model.\footnote{It is worth mentioning that this degeneracy is also shown on $\sigma_8$ for the cases of photometric and \hi-galaxy surveys (see Paper I), when one tries to fit mock data simulated assuming both density and RSDs, against spectra including density fluctuations only.}

To understand this, let us draw the reader's attention to the galaxy bias fiducial values of \autoref{tab:5bins}, chosen for the reference cosmology to produce the mock data, it is evident that these values are quite large. This is normal since the EMU survey as a radio continuum experiment probes very high redshifts, where the galaxy bias is expected to be rather large. In addition to this, we have already proved that an incomplete model chosen to fit the correct data can sometimes be insufficient to describe it successfully, leading to a misplaced/biased peak of the posterior. This, along with the fact that the galaxy bias extends to high values, leads the incomplete model to make erroneous overestimates of the galaxy bias nuisance parameters, which are counterbalanced by a rather low and therefore biased measurement on $\sigma_8$, which is of course imprinted on $S_8$ as well.

Despite this peculiar result for the incomplete model in the conservative scenario for the 5 bins, generally the biased estimates with the wrong model are those in the analysis with 2 very wide bins described in the previous subsection. This leads to the conclusion that the magnification contributing to the galaxy clustering is very significant, and it may not be neglected when wide redshift bins are chosen. This makes  sense, too, since the magnification bias of \autoref{eq:W_mag} is an integrated effect, implying that the wider the redshift range of the sources who are inside the bin, the more enhanced the effect of the magnification will be, leading to important biases when it is not considered.

By comparing the results with the 2-bin case, one can easily appreciate that the constraints obtained with the five narrower bins are tighter, especially on $h$. This can be attributed to the fact that the parameter's effect on the power spectrum can be determined through an accurate determination of  its redshift dependence, which is more precise with narrower redshift  bins.

The findings for the case of 5 Gaussian bins are quantitatively summarised in \autoref{tab:results_EMU5G_LCDM}.
\begin{table*}
\centering
\caption{Same as \autoref{tab:results_EMU2G_LCDM}, but for the case of 5 Gaussian bins.}
\begin{tabular}{lllcllcll}
    \hline
    \multicolumn{9}{c}{5 Gaussian bins ($\Lambda$CDM)} \\
    \hline
   & \multicolumn{2}{c}{Ideal scenario} && \multicolumn{2}{c}{Realistic scenario} && \multicolumn{2}{c}{Conservative scenario} \\
    \cline{2-3}\cline{5-6}\cline{8-9}
   & \multicolumn{1}{c}{den} & \multicolumn{1}{c}{den+mag} && \multicolumn{1}{c}{den} & \multicolumn{1}{c}{den+mag} && \multicolumn{1}{c}{den} & \multicolumn{1}{c}{den+mag} \\
    \hline
    \hline
    $S_8$ & $0.826\pm0.027$ & $0.827\pm0.026$ && $0.83\pm0.20$ & $0.818\pm0.099$ && $0.62\pm0.11$ & $0.75\pm0.11$ \\
    $h$ & $0.699\pm0.059$ & $0.684\pm0.057$ && $0.698\pm0.057$ & $0.680\pm0.055$ && $0.683\pm0.075$ & $0.669\pm0.072$ \\
    \hline
\end{tabular}
\label{tab:results_EMU5G_LCDM}
\end{table*}

\subsection{Constraints on dark energy}
\begin{figure*}
\centering
\includegraphics[width=0.45\textwidth,trim={3.6cm 9cm 4cm 7cm},clip]{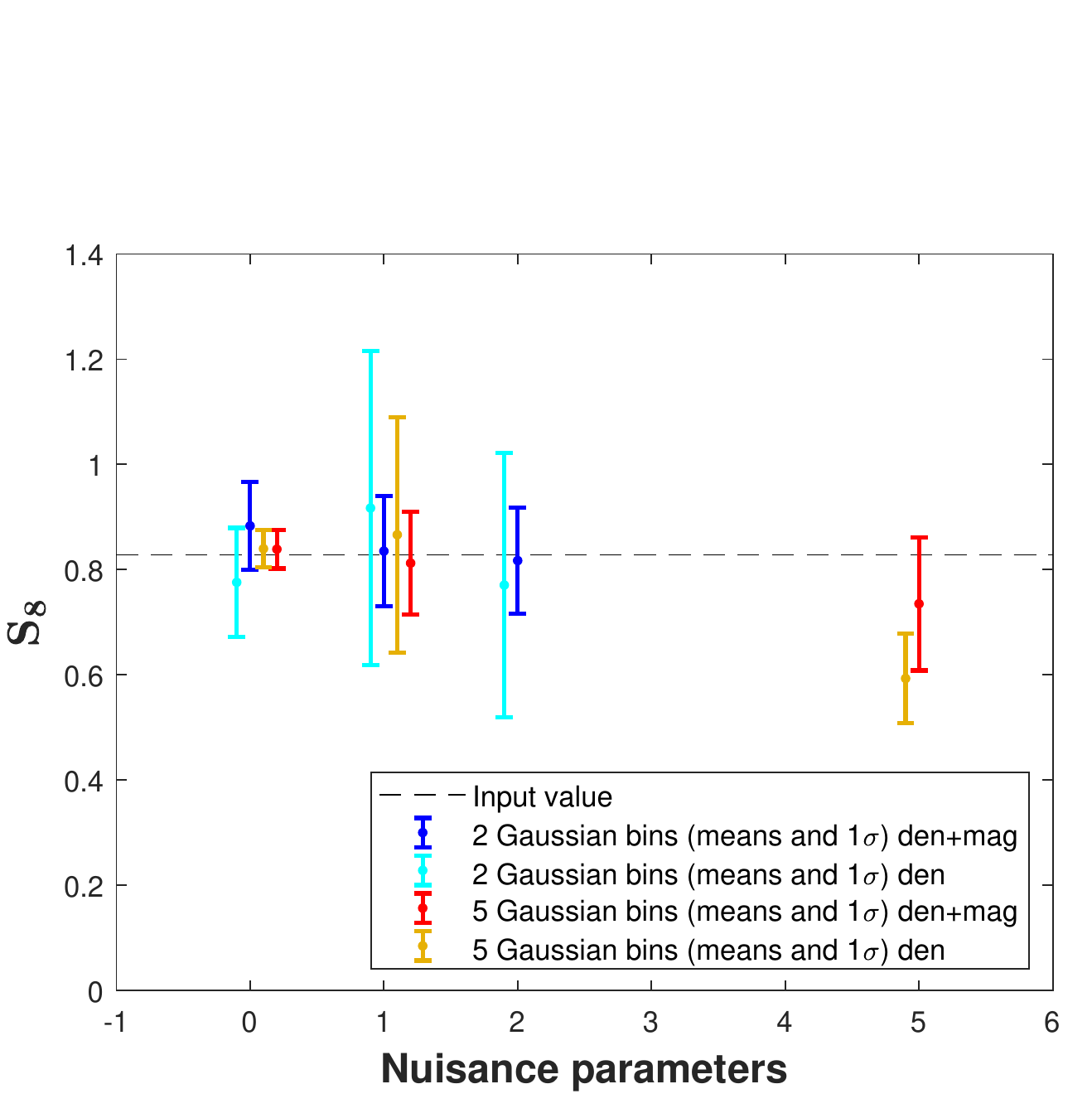}\includegraphics[width=0.45\textwidth,trim={3.6cm 9cm 4cm 7cm},clip]{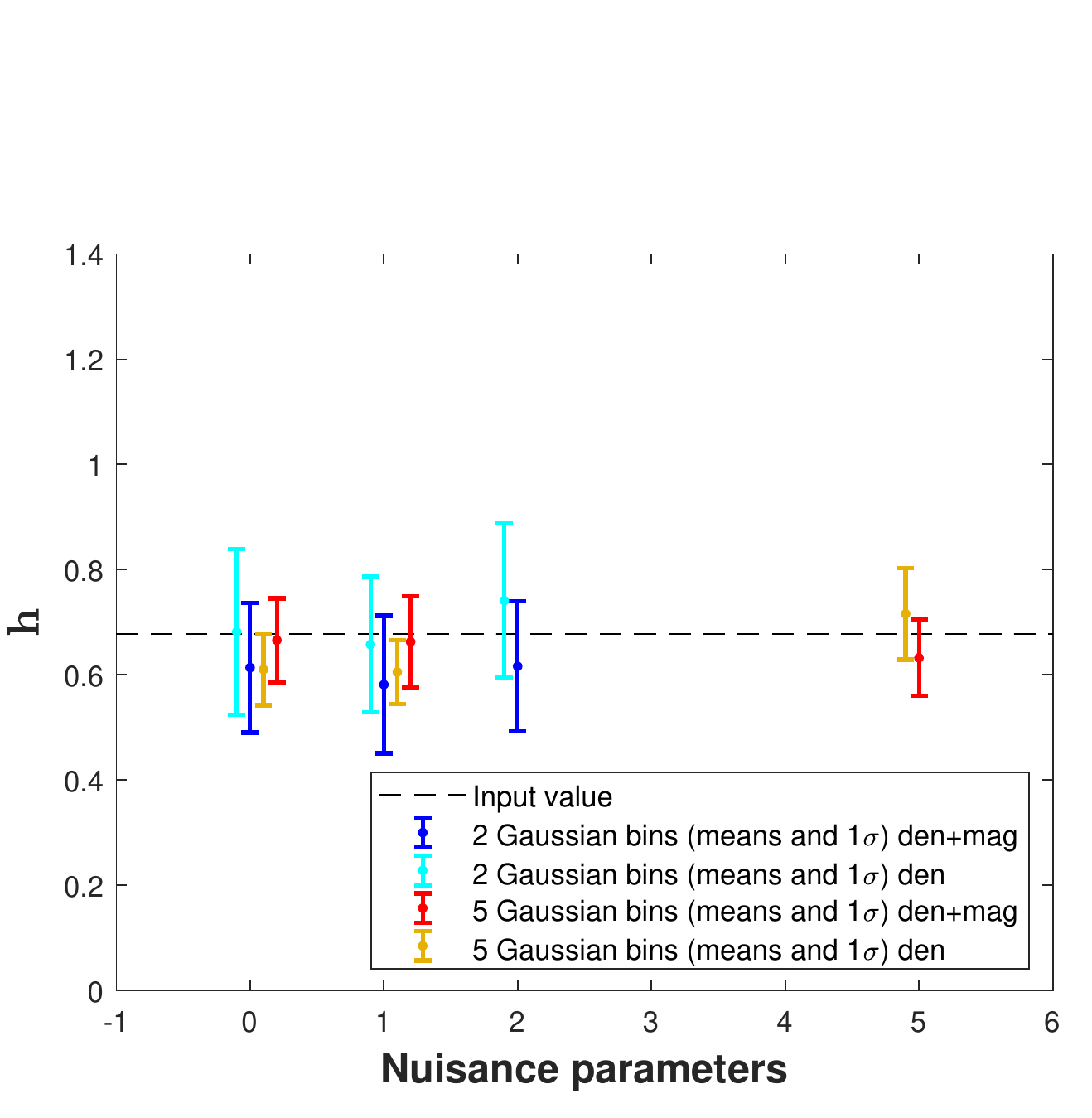}\\\includegraphics[width=0.45\textwidth,trim={3.6cm 9cm 4cm 9.5cm},clip]{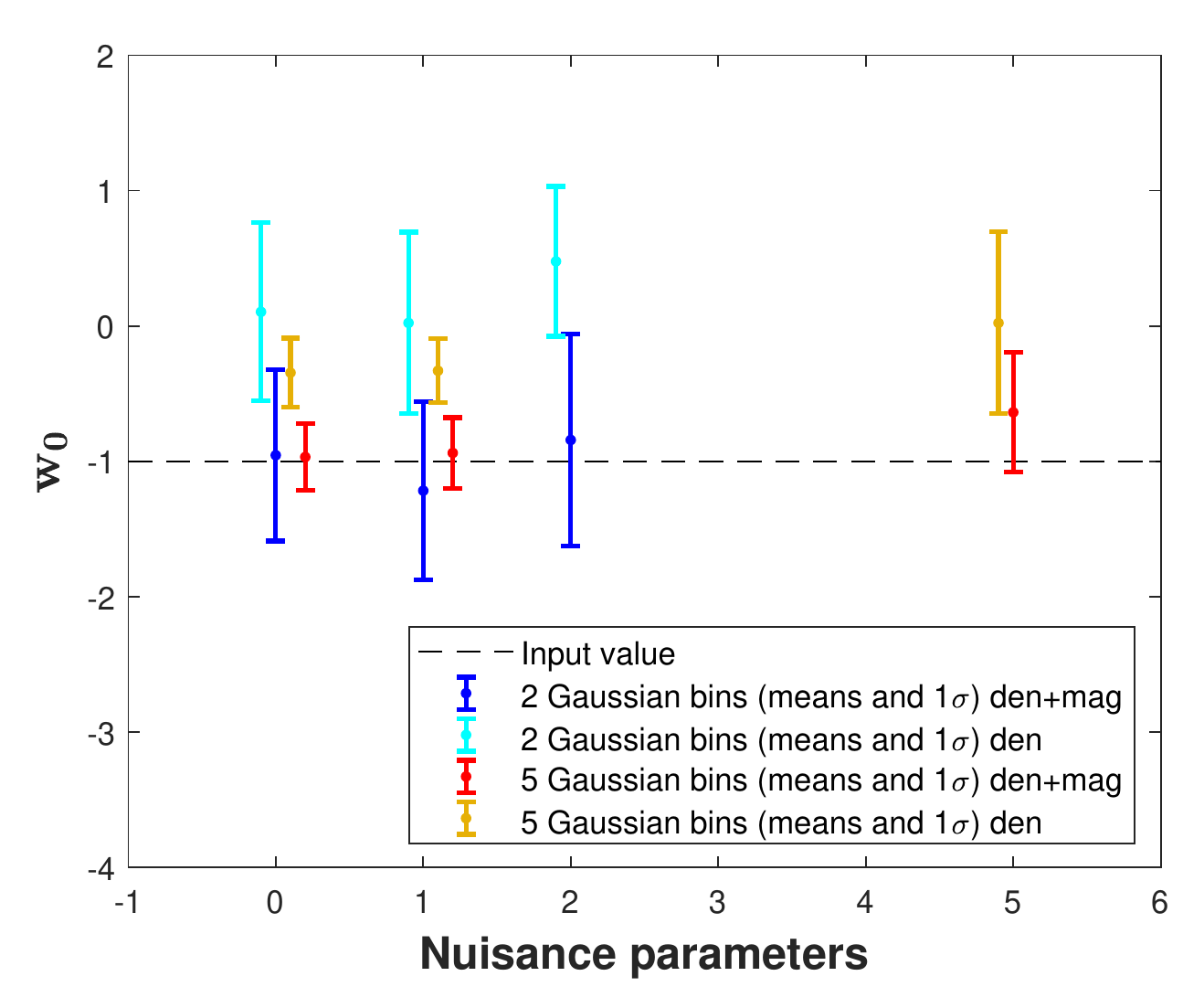}\includegraphics[width=0.45\textwidth,trim={3.6cm 9cm 4cm 9.5cm},clip]{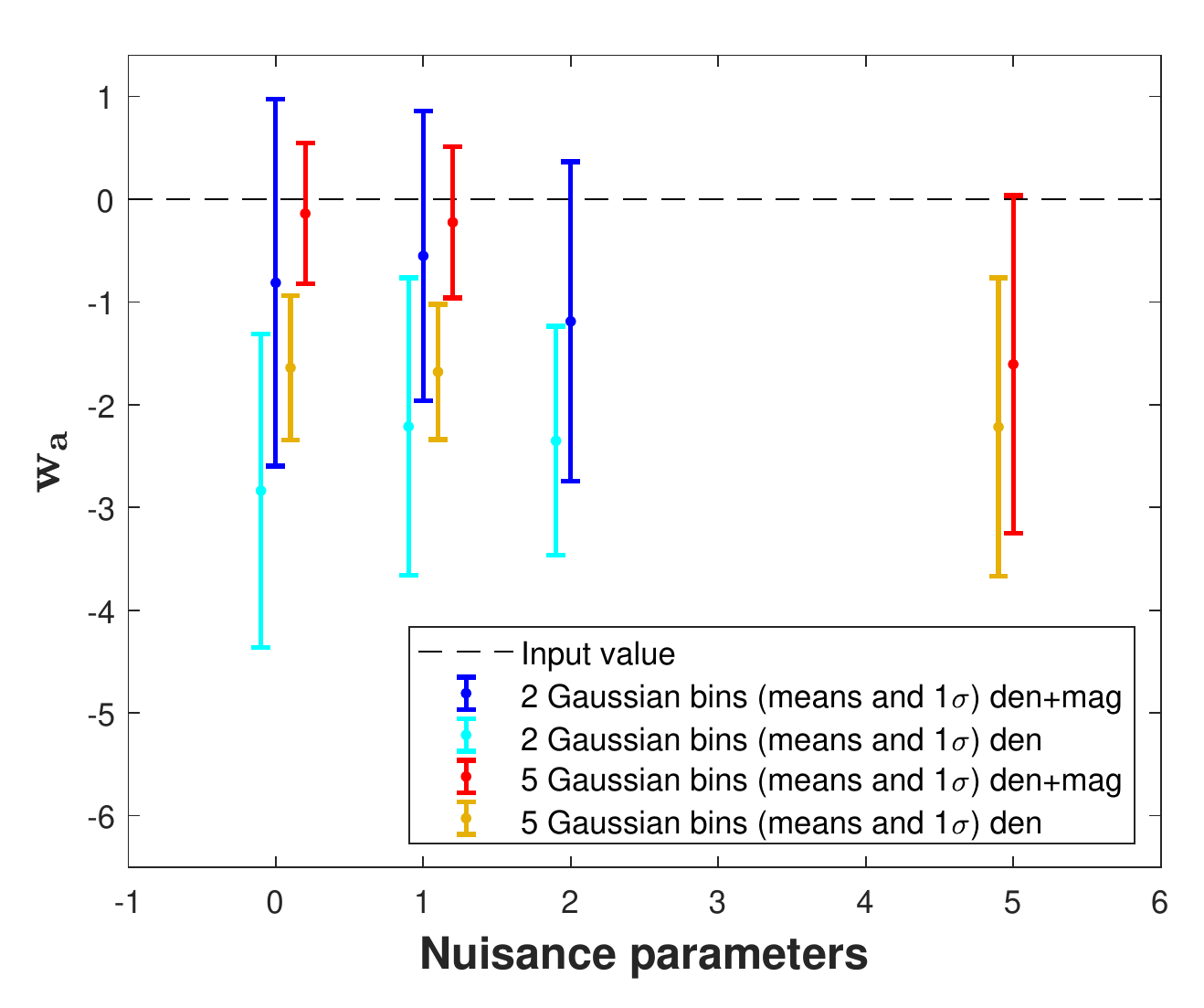}
\caption{Same as \autoref{fig:LCDM}, but for the dark energy parameter set.}
\label{fig:w0waCDM}
\end{figure*}

Let us know move to the first extension to \lcdm\ considered, namely dynamical dark energy as in \autoref{sec:DE}. The $68\%$ marginal confidence intervals and means on the cosmological set $\{S_8,\,h,\,w_0,\,w_a\}$ are presented in \autoref{fig:w0waCDM}, \autoref{tab:results_EMU2G_w0waCDM}, and \autoref{tab:results_EMU5G_w0waCDM}.
\begin{table*}
\centering
\caption{Same as \autoref{tab:results_EMU2G_LCDM}, but for dark energy.}
\begin{tabular}{lllcllcll}
    \hline
    \multicolumn{9}{c}{2 Gaussian bins (DE)} \\
    \hline
   & \multicolumn{2}{c}{Ideal scenario} && \multicolumn{2}{c}{Realistic scenario} && \multicolumn{2}{c}{Conservative scenario} \\
    \cline{2-3}\cline{5-6}\cline{8-9}
   & \multicolumn{1}{c}{den} & \multicolumn{1}{c}{den+mag} && \multicolumn{1}{c}{den} & \multicolumn{1}{c}{den+mag} && \multicolumn{1}{c}{den} & \multicolumn{1}{c}{den+mag} \\
    \hline
    \hline
    $S_8$ & $0.78\pm0.10$ & $0.883\pm0.083$ && $0.92\pm0.30$ & $0.84\pm0.10$ && $0.77\pm0.25$ & $0.82\pm0.10$ \\
    $h$ & $0.68\pm0.16$ & $0.61\pm0.13$ && $0.66\pm0.13$ & $0.58\pm0.13$ && $0.74\pm0.15$ & $0.66\pm0.12$ \\
        $w_0$ & $0.10\pm0.66$ & $-0.96\pm0.63$ && $0.02\pm0.67$ & $-1.22\pm0.66$ && $0.48\pm0.55$ & $-0.84\pm0.78$ \\
            $w_a$ & $-2.8\pm1.5$ & $-0.8\pm1.8$ && $-2.2\pm1.4$ & $-0.6\pm1.4$ && $-2.4\pm1.1$ & $-1.2\pm1.6$ \\
    \hline
\end{tabular}
\label{tab:results_EMU2G_w0waCDM}
\end{table*}
\begin{table*}
\centering
\caption{Same as \autoref{tab:results_EMU5G_LCDM}, but for dark energy.}
\begin{tabular}{lllcllcll}
    \hline
    \multicolumn{9}{c}{5 Gaussian bins (DE)} \\
    \hline
   & \multicolumn{2}{c}{Ideal scenario} && \multicolumn{2}{c}{Realistic scenario} && \multicolumn{2}{c}{Conservative scenario} \\
    \cline{2-3}\cline{5-6}\cline{8-9}
   & \multicolumn{1}{c}{den} & \multicolumn{1}{c}{den+mag} && \multicolumn{1}{c}{den} & \multicolumn{1}{c}{den+mag} && \multicolumn{1}{c}{den} & \multicolumn{1}{c}{den+mag} \\
    \hline
    \hline
    $S_8$ & $0.840\pm0.035$ & $0.839\pm0.037$ && $0.87\pm0.22$ & $0.812\pm0.098$ && $0.593\pm0.085$ & $0.73\pm0.13$ \\
    $h$ & $0.610\pm0.068$ & $0.666\pm0.080$ && $0.605\pm0.060$ & $0.663\pm0.086$ && $0.716\pm0.087$ & $0.632\pm0.072$ \\
        $w_0$ & $-0.35\pm0.26$ & $-0.97\pm0.25$ && $-0.33\pm0.24$ & $-0.94\pm0.26$ && $0.02\pm0.67$ & $-0.64\pm0.44$ \\
            $w_a$ & $-1.64\pm0.70$ & $-0.14\pm0.68$ && $-1.68\pm0.66$ & $-0.22\pm0.74$ && $-2.2\pm1.5$ & $-1.6\pm1.6$ \\    
    \hline
\end{tabular}
\label{tab:results_EMU5G_w0waCDM}
\end{table*}

Generally speaking, we find the same behaviour of constraints on $S_8$ and $h$ as for \lcdm, but there are a couple of points which nonetheless differ from the \lcdm\ results. The former is that in this parameterisation, the density-alone model for the 5 bins yields a slightly biased result on $h$ in the realistic scenario. The latter concerns that, in particular, the idealistic case constraints are a bit weaker than the \lcdm\ ones. This, of course, is due to the addition of the parameter set $\{w_0,\,w_a\}$, resulting in a larger statistical uncertainty in the posterior, keeping even the constraints for the wrong model and the 2 wide bins, consistent within 1$\sigma$ from the reference cosmology. Apart from that, regardless of the binning, a correct modelling yields comparable results for the realist and the conservative case, within $68\%$ from the fiducial values.

If we now focus on $\{w_0,\,w_a\}$, which constitutes one of the main points of our paper. It is evident that for any binning applied in the density-only model, since the reconstructed results are always biased on both parameters whether we introduce nuisance parameters to be marginalised over or not. In detail, we see that the picture of the analysis with the 2 bins is independent of the status of knowledge of the galaxy bias. The same is true for the 5 bins, apart form the conservative case where we get weakened results. It is worth noticing again that from the two configurations, the 5-bin choice yields better constraints. Indeed, after having a look at the mean values estimated by the incomplete model, we can really appreciate that the bias is more pronounced with the wider bins (cyan error bars compared to yellow ones). Generally, it is obvious that the correct model (blue in the 2-bin and red in the 5-bin case) always accepts the fiducial values $w_0=-1$ and $w_a=0$ within the $68\%$ marginal error.

Given these results, we infer that fitting the mock data with the complete model containing the same full information (density fluctuations and magnification) does not point to a spurious dark energy extension of the \lcdm\ model, which would not otherwise be the case if we ignored the magnification. This demonstrates the fact that the inclusion of the magnification bias on the galaxy density field is indispensable, in order to avoid misinterpretation of the results on the cosmological parameter estimation.

\subsection{Constraints on modified gravity}
\begin{figure*}
\centering
\includegraphics[width=0.45\textwidth,trim={3.6cm 9cm 4cm 7cm},clip]{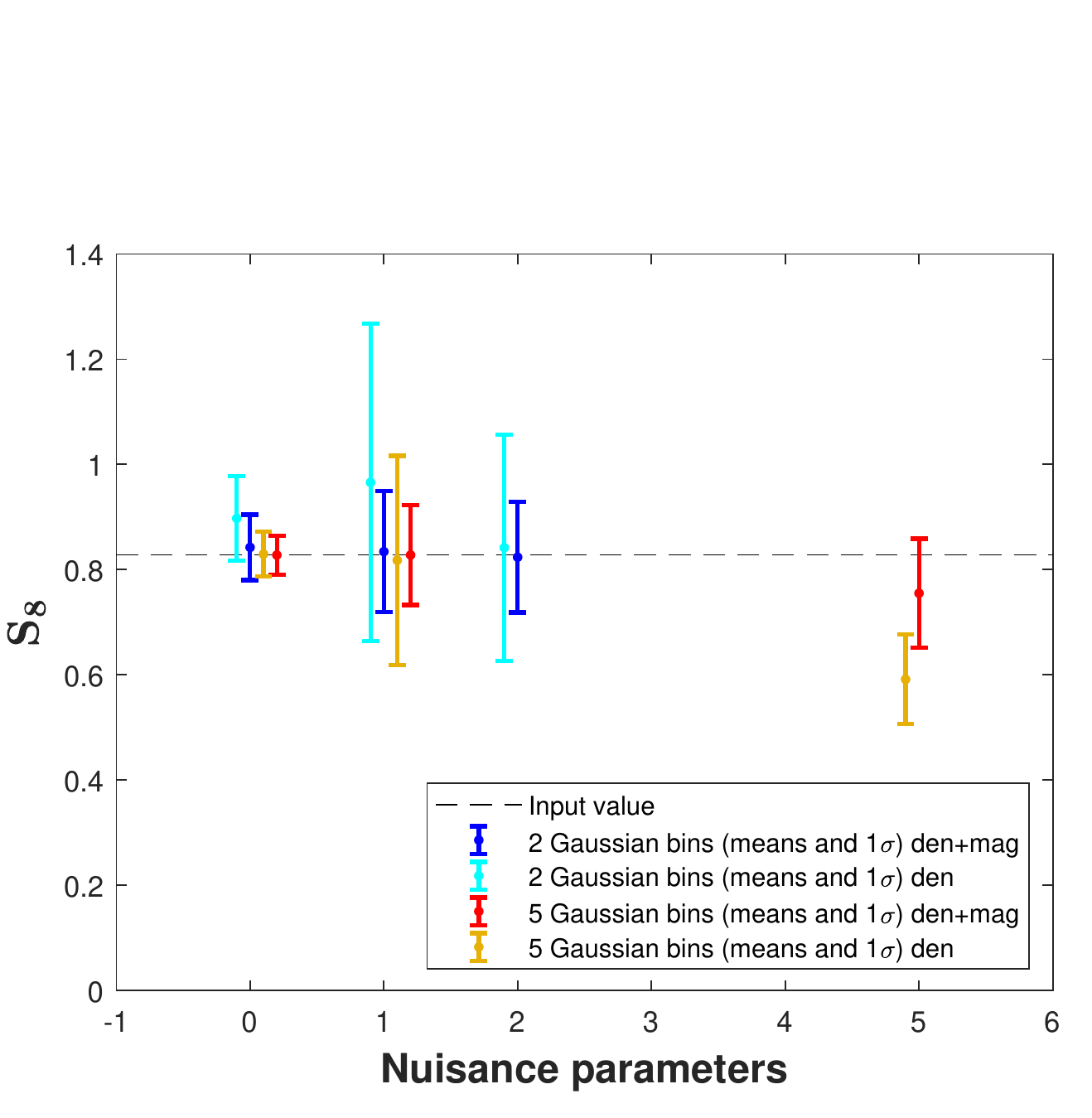}\includegraphics[width=0.45\textwidth,trim={3.6cm 9cm 4cm 7cm},clip]{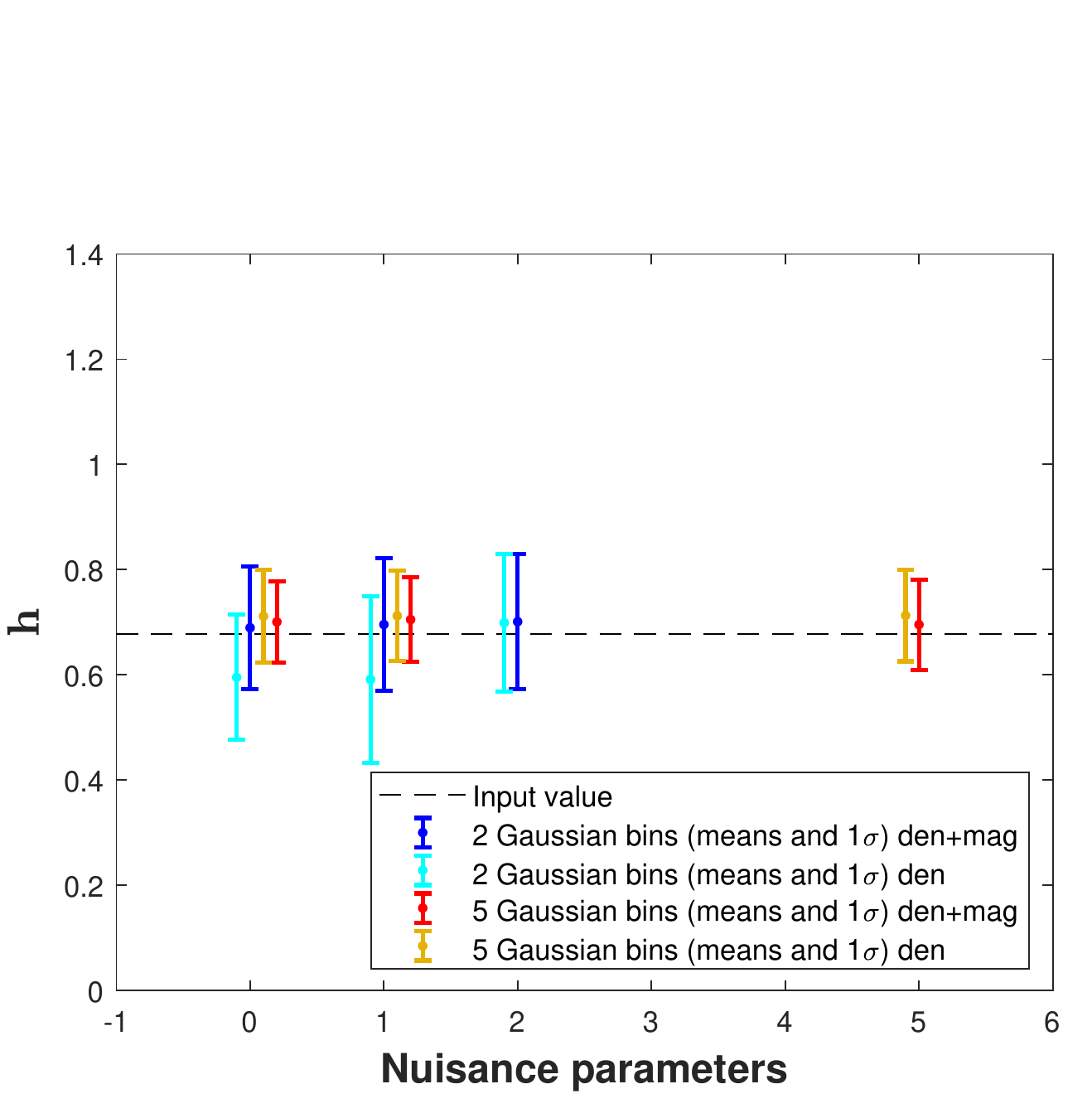}\\\includegraphics[width=0.45\textwidth,trim={3.6cm 9cm 4cm 9.5cm},clip]{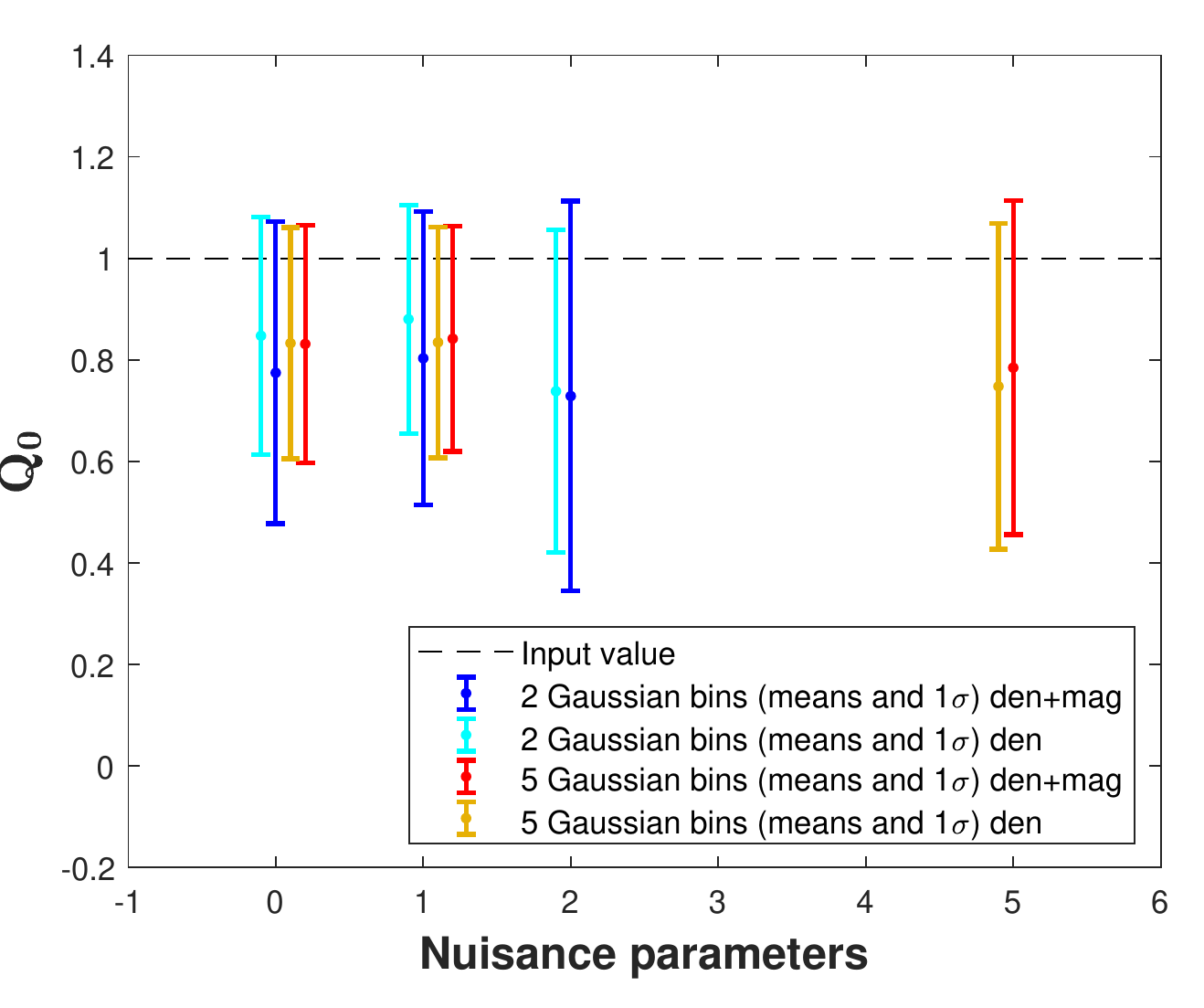}\includegraphics[width=0.45\textwidth,trim={3.6cm 9cm 4cm 9.5cm},clip]{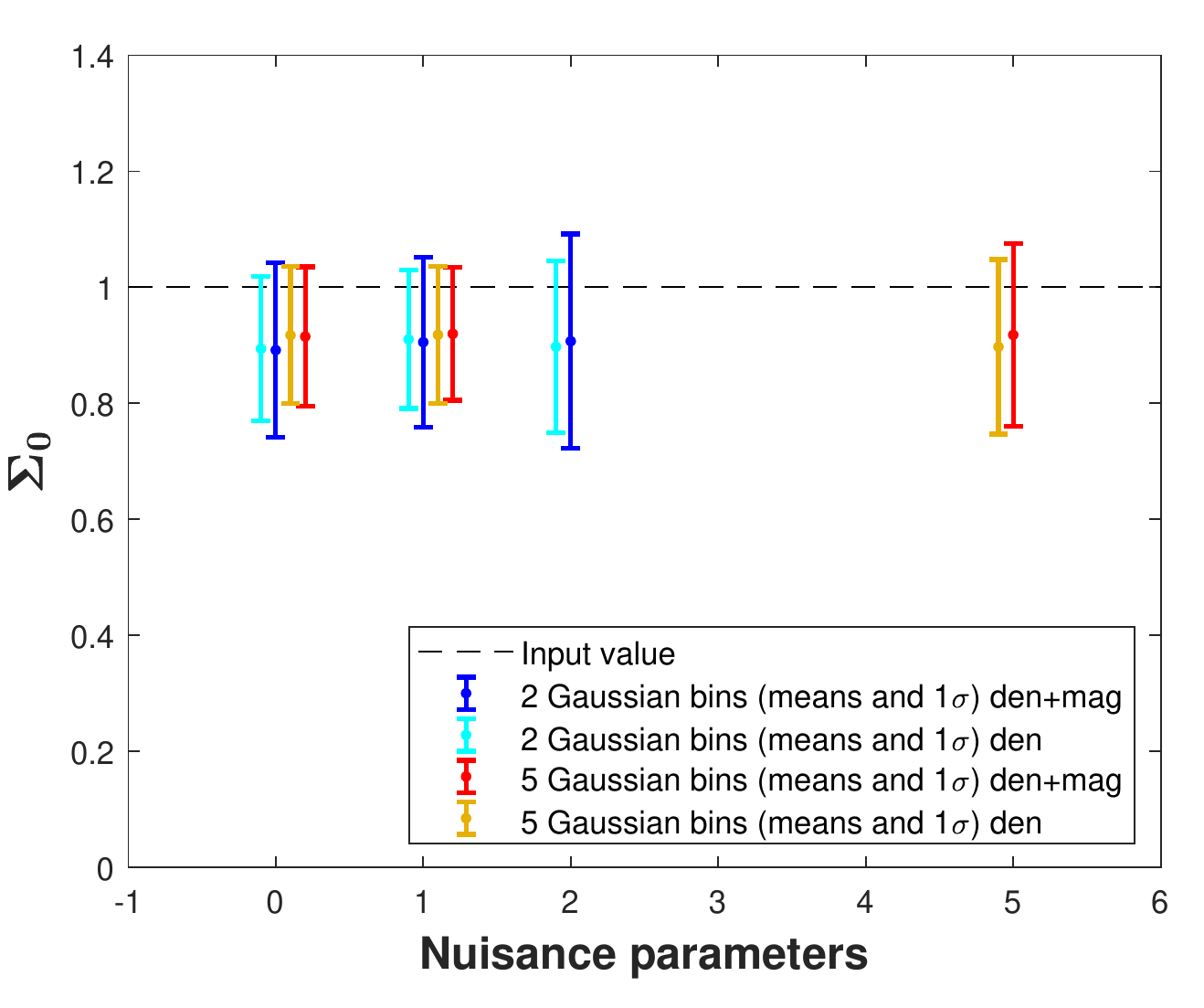}
\caption{Same as \autoref{fig:LCDM}, but for the modified gravity parameter set.}
\label{fig:QSCDM}
\end{figure*}

Finally, \autoref{fig:QSCDM}, \autoref{tab:results_EMU2G_Q0S0CDM}, and \autoref{tab:results_EMU5G_Q0S0CDM} present the parameter constraints on the modified gravity model parameters $\{S_8,\,h,\,Q_0,\,\Sigma_0\}$.
\begin{table*}
\centering
\caption{Same as \autoref{tab:results_EMU2G_LCDM}, but for modified gravity.}
\begin{tabular}{lllcllcll}
    \hline
    \multicolumn{9}{c}{2 Gaussian bins (MG)} \\
    \hline
   & \multicolumn{2}{c}{Ideal scenario} && \multicolumn{2}{c}{Realistic scenario} && \multicolumn{2}{c}{Conservative scenario} \\
    \cline{2-3}\cline{5-6}\cline{8-9}
   & \multicolumn{1}{c}{den} & \multicolumn{1}{c}{den+mag} && \multicolumn{1}{c}{den} & \multicolumn{1}{c}{den+mag} && \multicolumn{1}{c}{den} & \multicolumn{1}{c}{den+mag} \\
    \hline
    \hline
    $S_8$ & $0.897\pm0.080$ & $0.842\pm0.062$ && $0.97\pm0.30$ & $0.83\pm0.11$ && $0.84\pm0.22$ & $0.82\pm0.11$ \\
    $h$ & $0.60\pm0.12$ & $0.69\pm0.12$ && $0.59\pm0.16$ & $0.70\pm0.13$ && $0.70\pm0.13$ & $0.70\pm0.13$ \\
        $Q_0$ & $0.85\pm0.23$ & $0.77\pm0.30$ && $0.88\pm0.23$ & $0.80\pm0.29$ && $0.74\pm0.32$ & $0.73\pm0.38$ \\
            $\Sigma_0$ & $0.89\pm0.12$ & $0.89\pm0.15$ && $0.91\pm0.12$ & $0.91\pm0.15$ && $0.90\pm0.15$ & $0.91\pm0.19$ \\
    \hline
\end{tabular}
\label{tab:results_EMU2G_Q0S0CDM}
\end{table*}
\begin{table*}
\centering
\caption{Same as \autoref{tab:results_EMU5G_LCDM}, but for modified gravity.}
\begin{tabular}{lllcllcll}
    \hline
    \multicolumn{9}{c}{5 Gaussian bins (MG)} \\
    \hline
   & \multicolumn{2}{c}{Ideal scenario} && \multicolumn{2}{c}{Realistic scenario} && \multicolumn{2}{c}{Conservative scenario} \\
    \cline{2-3}\cline{5-6}\cline{8-9}
   & \multicolumn{1}{c}{den} & \multicolumn{1}{c}{den+mag} && \multicolumn{1}{c}{den} & \multicolumn{1}{c}{den+mag} && \multicolumn{1}{c}{den} & \multicolumn{1}{c}{den+mag} \\
    \hline
    \hline
    $S_8$ & $0.829\pm0.043$ & $0.828\pm0.037$ && $0.82\pm0.20$ & $0.83\pm0.095$ && $0.591\pm0.085$ & $0.75\pm0.10$ \\
    $h$ & $0.711\pm0.088$ & $0.700\pm0.077$ && $0.712\pm0.086$ & $0.705\pm0.080$ && $0.713\pm0.087$ & $0.700\pm0.086$ \\
        $Q_0$ & $0.83\pm0.23$ & $0.83\pm0.23$ && $0.83\pm0.23$ & $0.84\pm0.22$ && $0.75\pm0.32$ & $0.78\pm0.33$ \\
            $\Sigma_0$ & $0.93\pm0.12$ & $0.92\pm0.12$ && $0.92\pm0.12$ & $0.92\pm0.11$ && $0.90\pm0.15$ & $0.92\pm0.16$ \\    
    \hline
\end{tabular}
\label{tab:results_EMU5G_Q0S0CDM}
\end{table*}

We see that for the 2-bins and for both the wrong (cyan error bar) and the correct (blue error bar) model, the results on $\{S_8,\,h\}$ are always within $68\%$ from the fiducial values, and once again the same pattern follows, with the degeneracy on $S_8$ and its alleviation after magnification is added in the realistic and the conservative case, which again give comparable results. When it comes to the narrower 5 bins, we have a similar behaviour with the exception that the constraints are more stringent, and there is a biased underestimation of the $S_8$ with the incomplete model (yellow error bar) in the conservative case. Also, the constraining power here for both binning scenarios on the set $\{S_8,\,h\}$ is similar to the case of dark energy.

Concerning the modified gravity parameters $\{Q_0,\,\Sigma_0\}$, if any of these two parameters deviates from unity, this would indicate that the \lcdm\ model possibly needs to be replaced by a modified theory of gravity. Nonetheless, we can see for both binning configurations and both models that the results are comparable, while all the estimates are unbiased with respect to the fiducial input value. In addition, the narrower 5 redshift bins yield  slightly tighter constraints than the 2-bin case. 

Overall, we can conclude that even after ignoring the magnification correction in galaxy clustering, we are not able to see a biased result on the $\{Q_0,\,\Sigma_0\}$ that would, incorrectly of course, imply that the vanilla \lcdm\ model is not the complete theory to describe the mock data.% The main reason for this is probably the fact that in our analysis, after implementing the Limber approximation, we exclude the ultra large scales which are considered to be sensitive to testing alternative theories of gravity, beyond the standard \lcdm\ theory. 

\subsection{Including redshift-space distortions}
\label{sec:RSD}
At last, we examine the impact of RSDs in the analysis. In Paper I, we have already presented results that show, for optical/near-IR and radio \hi-line galaxy surveys, that if one neglects RSDs when fitting against the data, one can induce biases in the cosmological parameter estimation.

In this case,  we create the mock data including all terms in \autoref{eq:W_tot}. We focus on the idealistic scenario, where the galaxy bias is perfectly known, as if no deviation from the results described above is found in this case, we even less expect to see any for the realistic and conservative cases. We constrain the parameter set $\{\om,\,h,\sigma_8\}$ with four different constructions of the theory vector: $i)$ density only; $ii)$ density and magnification (these two corresponding to what discussed in the previous subsections); $iii)$ density and RSDs; $iv)$ and density, RSDs, and magnification.

\begin{figure*}
\centering
\includegraphics[width=0.45\textwidth,trim={4cm 9cm 4cm 7cm},clip]{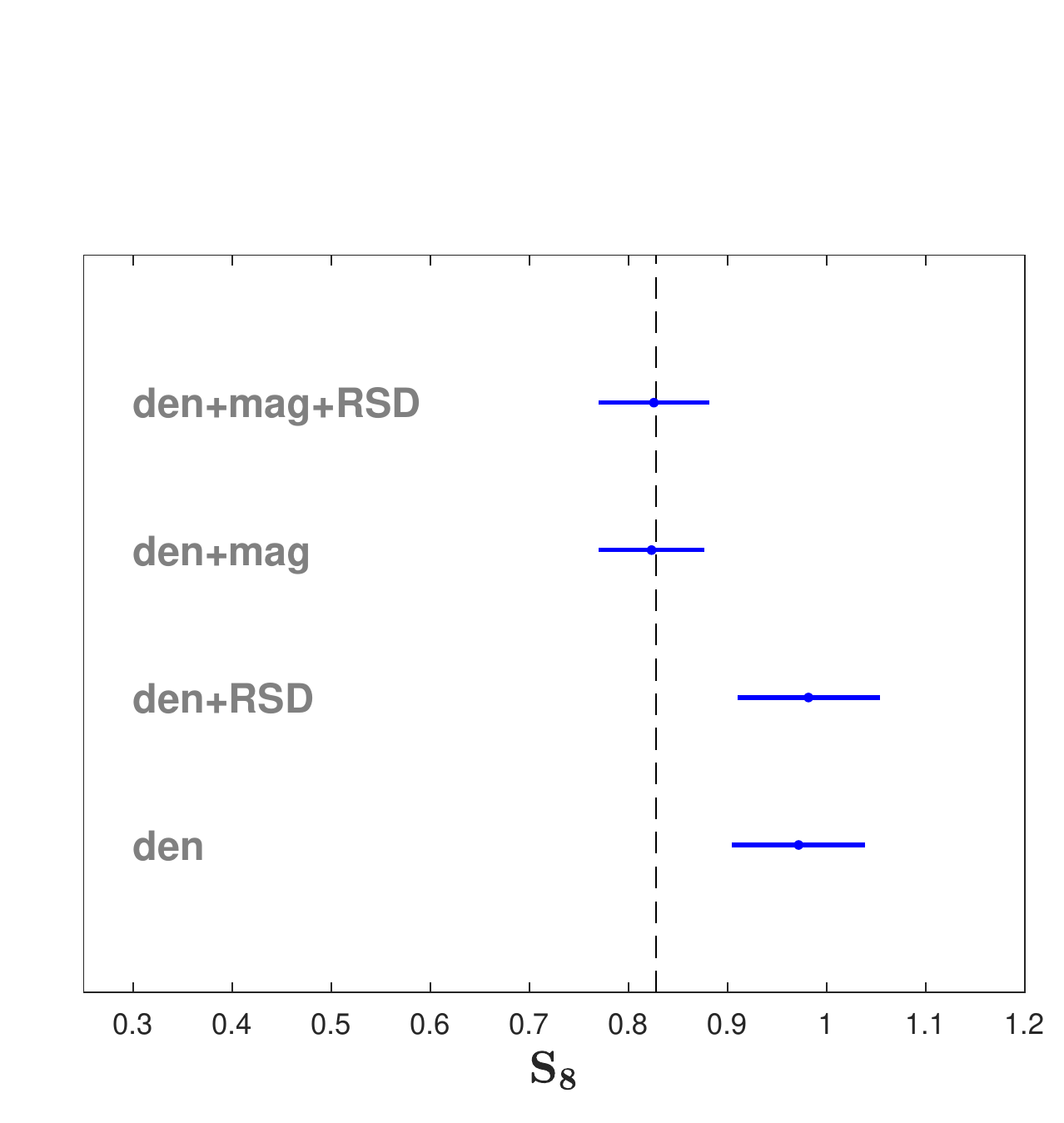}\includegraphics[width=0.45\textwidth,trim={4cm 9cm 4cm 7cm},clip]{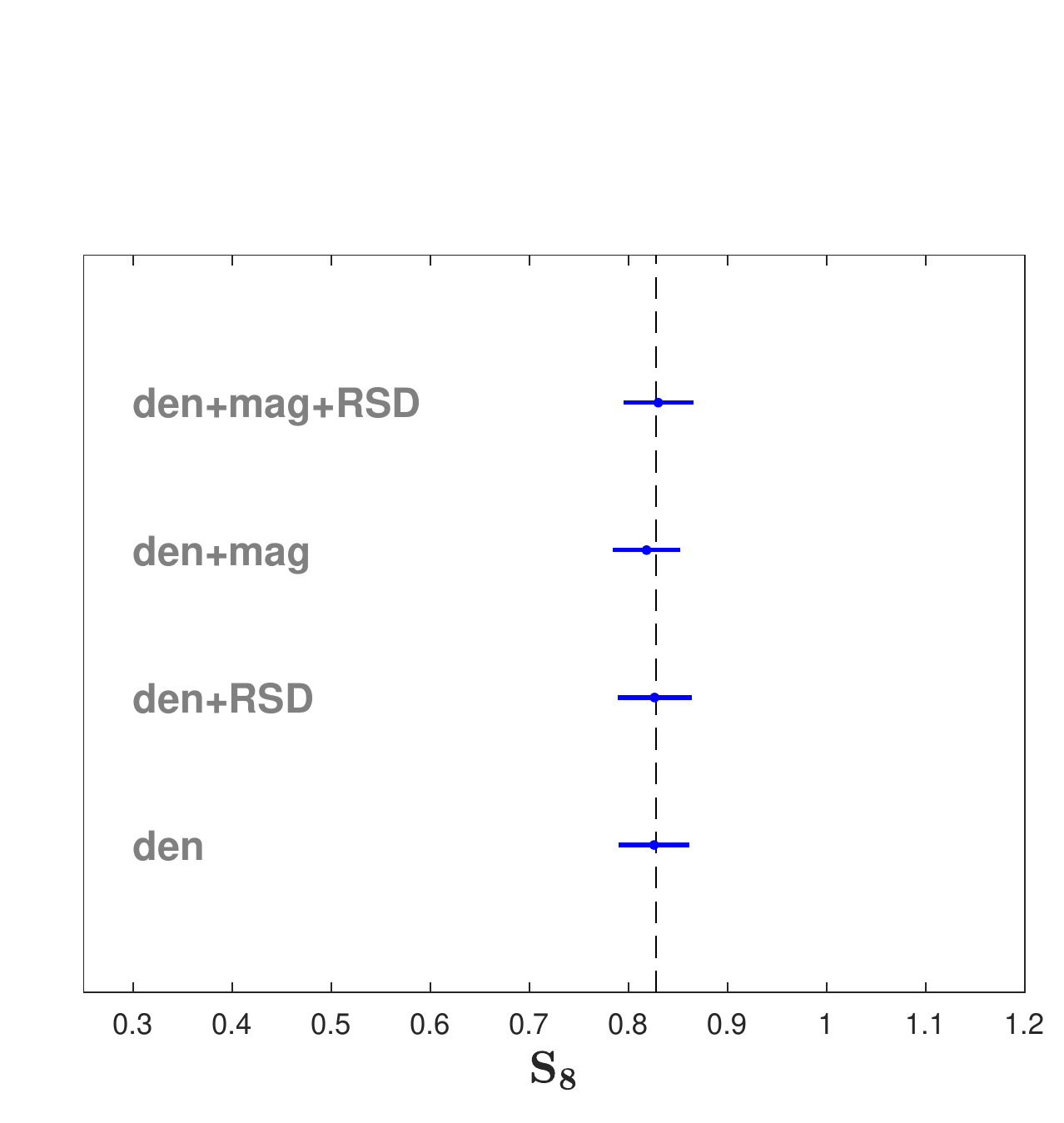}\\\includegraphics[width=0.45\textwidth,trim={4cm 9cm 4cm 9.5cm},clip]{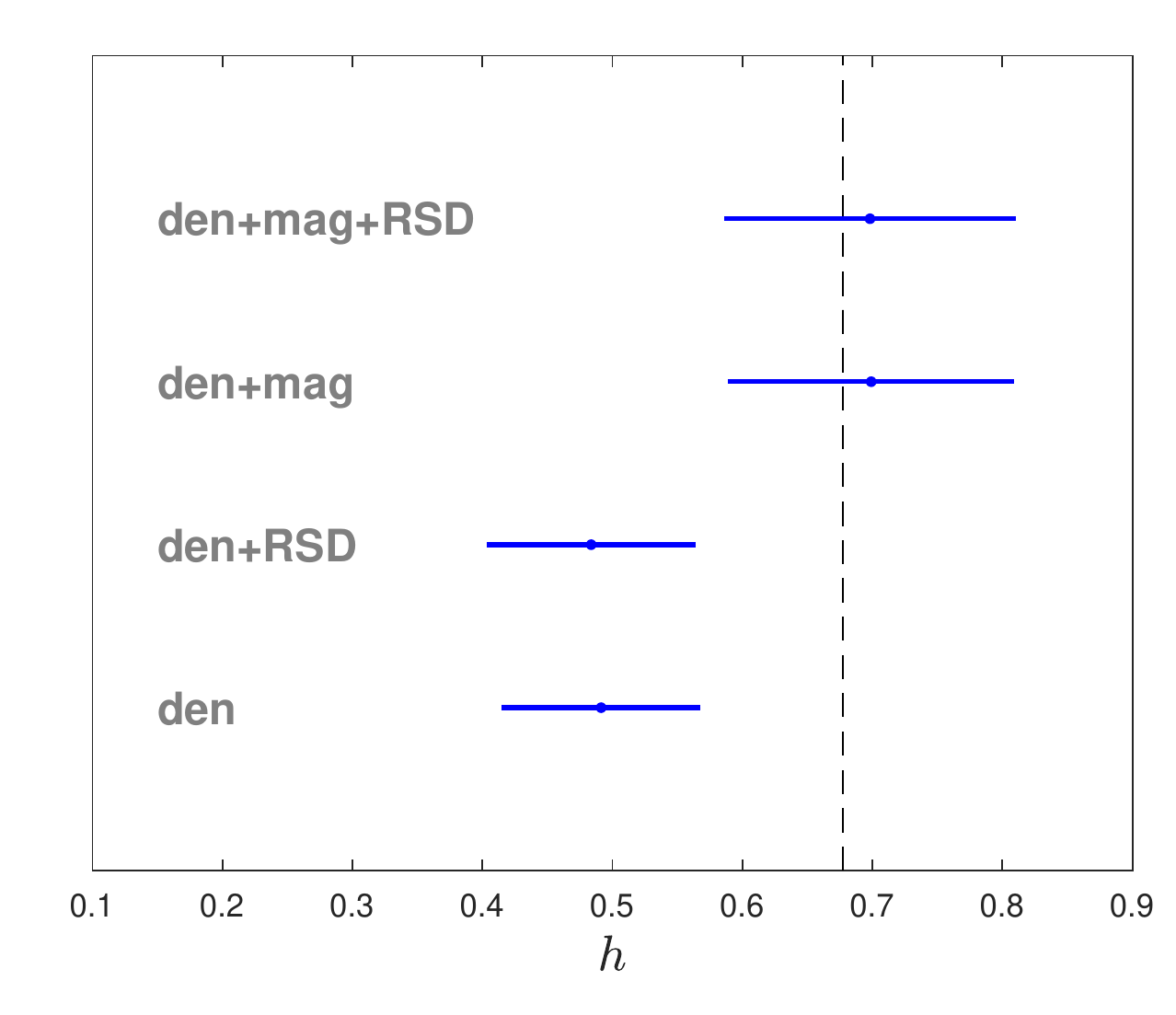}\includegraphics[width=0.45\textwidth,trim={4cm 9cm 4cm 9.5cm},clip]{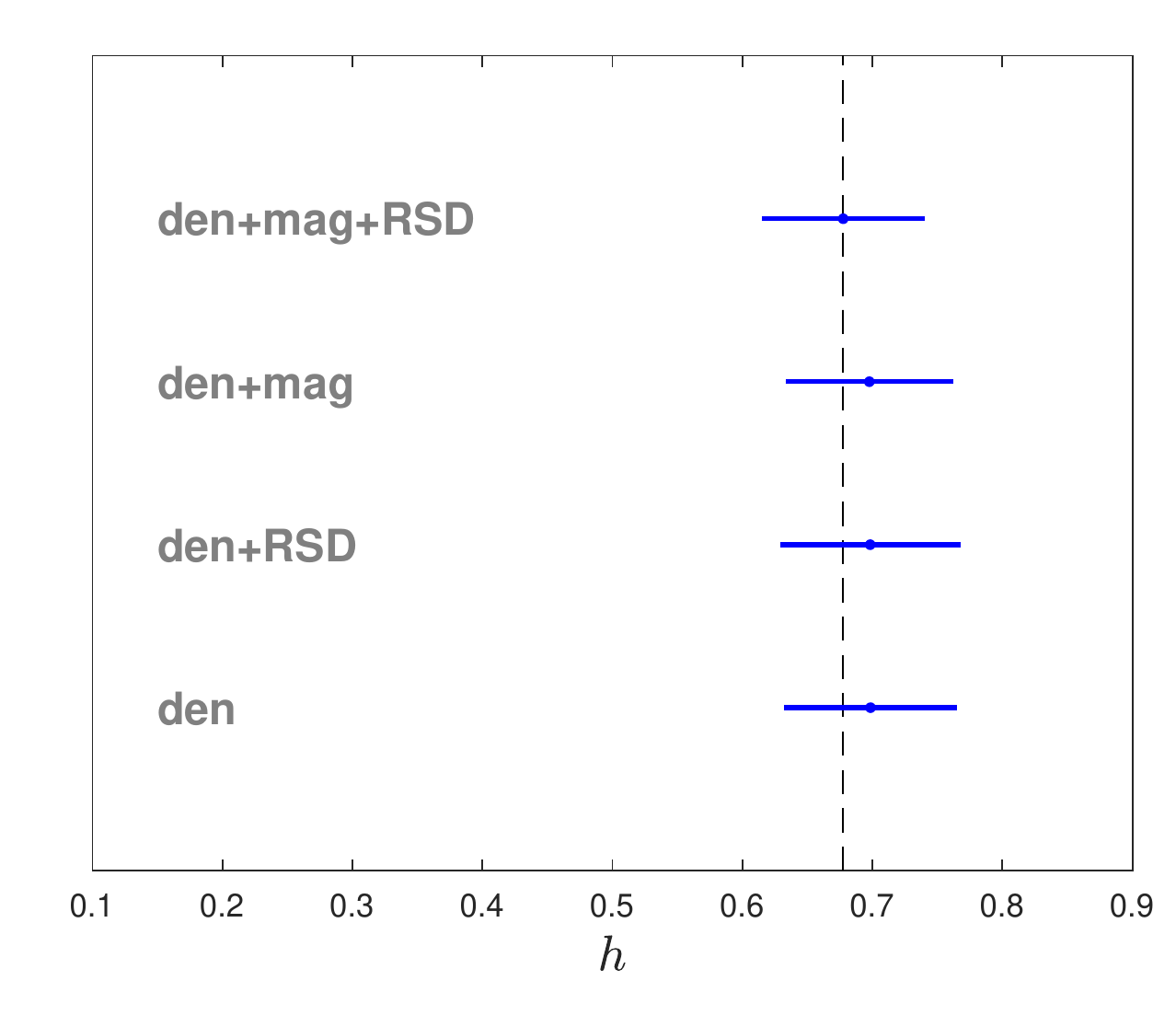}
\caption{EMU mean and 68\% constraints on the derived $S_8$ (top) and $h$ (bottom), cosmological parameter for 2 (left) and 5 (right) Gaussian bins in a \lcdm\ model where the galaxy bias is known exactly. Note that the data to be fitted are constructed incorporating both RSDs and the magnification bias correction on the galaxy density field in a \lcdm\ fiducial cosmology (vertical dashed line).}
\label{fig:RSD}
\end{figure*}

 \autoref{fig:RSD} presents the results for the four different models considered. The left panels show the constraints on the set $\{S_8,\,h\}$ for the 2-bin case. It is clear that there are biased estimates when the theory model includes the density fluctuations alone or the density along with the RSDs correction, neglecting in both cases the magnification bias. On the contrary, the theory model that contains the full information (density, RSDs and magnification) as the mock data is well within $68\%$ from the reference fiducial values, and so does the model which considers the density field and the magnification flux, but ignoring now RSDs. As for the results of the 5-bin case shown in the right panels, it is obvious that the constraints are better on both $S_8$ and $h$, while there are no biased estimates at all with any of the three incomplete models tested.
%\begin{table*}
%\centering
%\caption{Means and corresponding $68\%$ marginal error intervals on cosmological parameters for the EMU radio continuum galaxy survey applying 2 Gaussian bins with the \lcdm\ model. Four theory vector constructions are tested by fitting the mock data created with the full clustering information (density+RSDs+magnification).}
%\begin{tabular}{lllcll}
%    \hline
%    \multicolumn{6}{c}{2 Gaussian bins ($\Lambda$CDM)}\\
%    \cline{1-6}
%   & \multicolumn{1}{c}{den} & \multicolumn{1}{c}{den+RSD} && \multicolumn{1}{c}{den+mag} & \multicolumn{1}{c}{den+RSD+mag} \\
%    \hline
%    \hline
%    $S_8$ & $0.9718\pm0.0587$ & $0.9818\pm0.0631$ && $0.8233\pm0.0444$ & $0.8257\pm0.0474$ \\
%    $h$ & $0.4913\pm0.0691$ & $0.4836\pm0.0728$ && $0.6990\pm0.1028$ & $0.6981\pm0.1049$ \\
%    \hline
%\end{tabular}
%\label{tab:results_EMU2G_LCDM_RSD}
%\end{table*}
%\begin{table*}
%\centering
%\caption{Same as \autoref{tab:results_EMU2G_LCDM_RSD}, but for 5 Gaussian bins}
%\begin{tabular}{lllcll}
%    \hline
%    \multicolumn{6}{c}{5 Gaussian bins ($\Lambda$CDM)}\\
%    \cline{1-6}
%   & \multicolumn{1}{c}{den} & \multicolumn{1}{c}{den+RSD} && \multicolumn{1}{c}{den+mag} & \multicolumn{1}{c}{den+RSD+mag} \\
%    \hline
%    \hline
%    $S_8$ & $0.8259\pm0.0268$ & $0.8264\pm0.0284$ && $0.8183\pm0.0255$ & $0.8302\pm0.0265$ \\
%    $h$ & $0.6985\pm0.0588$ & $0.6983\pm0.0619$ && $0.6976\pm0.0571$ & $0.6775\pm0.0552$ \\
%    \hline
%\end{tabular}
%\label{tab:results_EMU5G_LCDM_RSD}
%\end{table*}

The above results, lead to the conclusion that the inclusion or the ignorance of the RSDs correcting term on the galaxy number counts, cannot affect our analysis at any extend, and can be safely ignored in our study. The reason behind this is the very large width of the redshift bins. Even when subdividing the redshift galaxy distribution into 5 bins, they are still quite wide in the redshift space, thus leading to a washing out of the RSD effect. Oppositely, narrower bins call for the inclusion of RSDs (see Paper I). On the other hand, this test provides a further confirmation that in the case of radio continuum surveys like EMU, the magnification bias ought to be included in the modelling, in order to avoid potential biases in the cosmological parameter estimation.

\section{Conclusions}
\label{sec:conclusion}
In the current work we have aimed to assess the effect of correctly including the weak lensing effect of magnification bias in galaxy number counts in a fully likelihood-based parameter estimation analysis. We have not only investigate standard \lcdm\ parameters, as well compelling extensions such as dynamical dark energy and a phenomenological parameterisation to modified gravity. To maximise the impact of magnification---which, being lensing, is an integrated effect---we have focussed the analysis on the specifications of deep radio continuum surveys using the Evolutionary Map of the Universe as a reference, for which we chose both two (very wide) and five (narrower yet broad) redshift bins. Moreover, we have restricted the harmonic-space angular power spectrum to the Limber approximation and the linear scales, and according to that, applied cuts on the multipole range. Then, we have created mock data including the magnification in the galaxy clustering and fit it with two theory vector constructions: one correctly including magnification bias, and another neglecting lensing.

In addition to that, we introduced a number of scenarios regarding the knowledge we have on the galaxy bias:
\begin{enumerate}
    \item An idealistic scenario where the galaxy bias is perfectly known;
    \item A pessimistic scenario where a free normalisation galaxy bias parameter is introduced at the whole redshift range;
    \item A conservative scenario that allows for a nuisance galaxy bias parameter for each bin.
\end{enumerate}

Considering all these cases, we summarise here the most important results obtained with the different cosmological models:
\begin{itemize}
\item[] \textit{$\varLambda$CDM} -- Here, the results we obtained with both binning configurations (Gaussian and top-hat) are comparable since the bins are always wide enough. In detail, we saw that there are biased estimates for the parameters $\{S_8,\,h\}$ when the galaxy bias is know exactly and if we neglect the magnification effect. This bias is not seen when we include nuisance parameters, but it is evident that the wrong theoretical model yields unconstrained results on the normalisation of the power spectrum $\sigma_8$ which is degenerate with galaxy bias. We lift this bias when we consider the magnification flux which is independent on the galaxy bias. Another point is that when the narrower binning is chosen, the parameters are more constrained due to the better redshift precision on the power spectrum. In addition to that, we appreciate in this case that there is a biased measurement in the conservative case with the incomplete model on $S_8$ owing to the overestimate of the nuisance galaxy bias parameters. This is also true for the following cosmological models that we examined. The results from now on were obtained adopting the more realistic case for the Gaussian redshift bins.
\item[] \textit{DE} -- Regarding the constraints on this CPL Dark Energy model, the biased estimates are not seen when we include the wrong theory vector in the 2-bin case, except the biased result on $h$ in the pessimistic scenario with the 5-bins. Overall, as in the \lcdm\ model, there are better constraints with the narrow binning over the wide one, and also degeneracy on $S_8$ which is alleviated with the magnification flux in the pessimistic and the conservative scenarios. As for the results on $\{w_0,\,w_a\}$, in the all the cases and the scenarios considered, the estimates with the incomplete model are biased. In the wide binning, the bias is slightly more enhanced since the magnification flux as a lensing effect becomes more important.   
\item[] \textit{MG} -- When we examine the Modified Gravity model, the results on $\{S_8,\,h\}$ are similar to those of the CPL, but with the only bias now seen only for the 5-bin conservative case on $S_8$. There are no biases on any parameter out of the set $\{Q_0,\,\Sigma_0\}$.%, probably due to the multipole cut applied on the very large scales which are not included and could provide valuable information on extensions to general relativity. 
\end{itemize}
In the final test we considered, we proved that the inclusion of the RSD correction in the galaxy clustering is not important in the case of radio continuum surveys like EMU, since the very poor redshift knowledge leads to the dilution of the effect. 

All the above results stress the importance that for the radio continuum surveys, the incorporation of the magnification flux is is necessary on the one hand, to avoid biases on the estimated parameters, and on the other hand, to break the degenerate relation between $\sigma_8$ and the galaxy bias. Also these biased estimates tend to increase when very wide bins are considered, a results that demonstrates the fact that the magnification effect becomes more important with time.

\section*{Acknowledgements}

KT and SC acknowledge support from the `Departments of Excellence 2018-2022' Grant (L.\ 232/2016) awarded by the Italian Ministry of Education, University and Research (\textsc{miur}). SC is funded by \textsc{miur} through Rita Levi Montalcini project `\textsc{prometheus} -- Probing and Relating Observables with Multi-wavelength Experiments To Help Enlightening the Universe's Structure'.

%%%%%%%%%%%%%%%%%%%%%%%%%%%%%%%%%%%%%%%%%%%%%%%%%%



\bibliographystyle{mnras}
\bibliography{paper2ver1} % if your bibtex file is called example.bib

%\begin{thebibliography}{99}

%\end{thebibliography}

%%%%%%%%%%%%%%%%%%%%%%%%%%%%%%%%%%%%%%%%%%%%%%%%%%

%%%%%%%%%%%%%%%%% APPENDICES %%%%%%%%%%%%%%%%%%%%%
%\onecolumn

\appendix

\section{Top-hat bins}
\label{Top-hat bins}
 
\begin{figure*}
\centering

\includegraphics[width=0.45\textwidth,trim={3.6cm 9cm 4cm 7cm},clip]{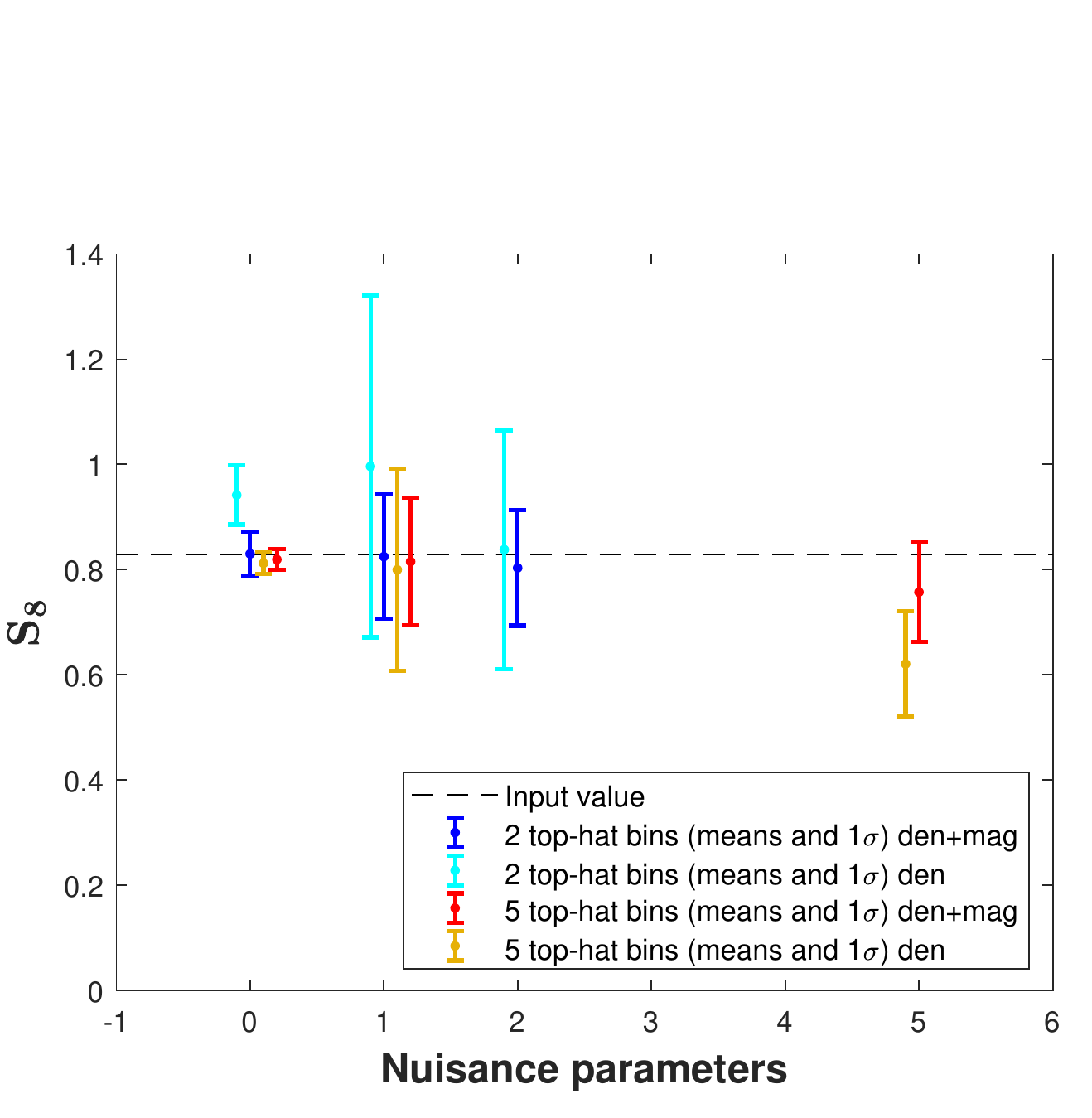}\includegraphics[width=0.45\textwidth,trim={3.6cm 9cm 4cm 7cm},clip]{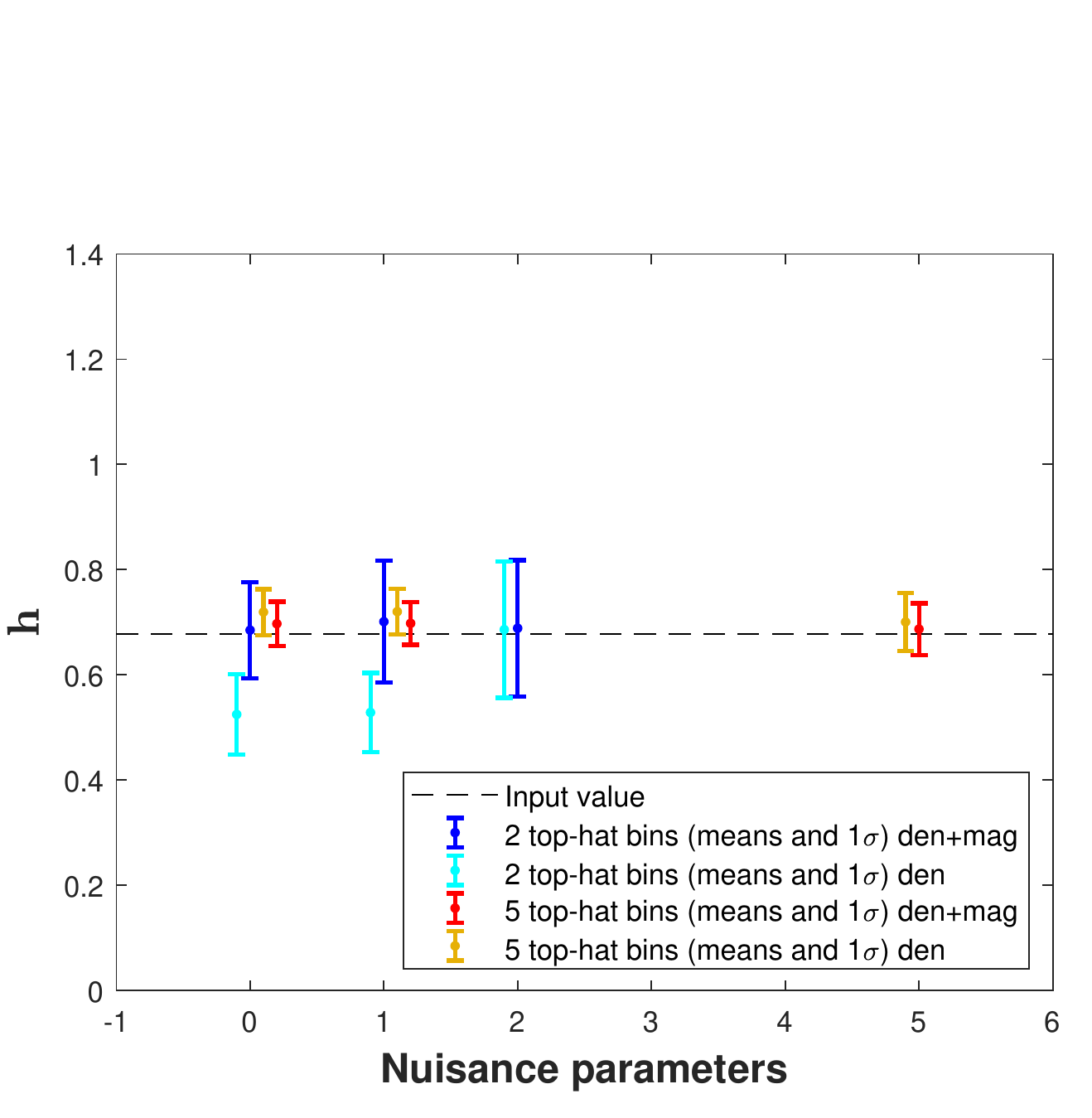}\\
\caption{EMU mean and 68\% constraints on the derived $S_8$ (left) and $h$ (right) cosmological parameter in top-hat (top) and Gaussian (bottom) bins as a function of the number of nuisance parameters for the \lcdm\ model. Note the different colours accounting for the number of bins and the density w/o magnification spectra fitting}
\label{fig:LCDM_TH}
\end{figure*}

Here, we present in ~\autoref{fig:LCDM_TH}, ~\autoref{tab:results_EMU2TH_LCDM} and ~\autoref{tab:results_EMU5TH_LCDM} the means and their corresponding $1\sigma$ confidence levels on the cosmological set $\{S_8,h\}$ in the case of 2 wide and 5 narrower top-hat bins for the ideal, the pessimistic and the conservative scenario.

\begin{table*}
\centering
\caption{Means and corresponding $68\%$ marginal error intervals on cosmological parameters for the EMU radio continuum galaxy survey applying 2 top-hat bins with the \lcdm\ model.}
\begin{tabular}{lllcllcll}
    \hline
    \multicolumn{9}{c}{2 top-hat bins ($\Lambda$CDM)} \\
    \hline
   & \multicolumn{2}{c}{Ideal scenario} && \multicolumn{2}{c}{Realistic scenario} && \multicolumn{2}{c}{Conservative scenario} \\
    \cline{2-3}\cline{5-6}\cline{8-9}
   & \multicolumn{1}{c}{den} & \multicolumn{1}{c}{den+mag} && \multicolumn{1}{c}{den} & \multicolumn{1}{c}{den+mag} && \multicolumn{1}{c}{den} & \multicolumn{1}{c}{den+mag} \\
    \hline
    \hline
    $S_8$ & $0.9415\pm0.0560$ & $0.8298\pm0.0422$ && $0.9960\pm0.3248$ & $0.8245\pm0.1185$ && $0.8379\pm0.2269$ & $0.8031\pm0.1101$ \\
    $h$ & $0.5244\pm0.0763$ & $0.6847\pm0.0915$ && $0.5281\pm0.07530$ & $0.7008\pm0.1161$ && $0.6857\pm0.1296$ & $0.6883\pm0.1296$ \\
    \hline
\end{tabular}
\label{tab:results_EMU2TH_LCDM}
\end{table*}

\begin{table*}
\centering
\caption{Means and corresponding $68\%$ marginal error intervals on cosmological parameters for the EMU radio continuum galaxy survey applying 5 top-hat bins with the \lcdm\ model.}
\begin{tabular}{lllcllcll}
    \hline
    \multicolumn{9}{c}{5 top-hat bins ($\Lambda$CDM)} \\
    \hline
   & \multicolumn{2}{c}{Ideal scenario} && \multicolumn{2}{c}{Realistic scenario} && \multicolumn{2}{c}{Conservative scenario} \\
    \cline{2-3}\cline{5-6}\cline{8-9}
   & \multicolumn{1}{c}{den} & \multicolumn{1}{c}{den+mag} && \multicolumn{1}{c}{den} & \multicolumn{1}{c}{den+mag} && \multicolumn{1}{c}{den} & \multicolumn{1}{c}{den+mag} \\
    \hline
    \hline
    $S_8$ & $0.8119\pm0.0205$ & $0.8191\pm0.0195$ && $0.7996\pm0.1921$ & $0.8150\pm0.1216$ && $0.6204\pm0.1003$ & $0.7570\pm0.0950$ \\
    $h$ & $0.7191\pm0.0435$ & $0.6969\pm0.0419$ && $0.7199\pm0.0436$ & $0.6976\pm0.0408$ && $0.7003\pm0.0547$ & $0.6867\pm0.0491$ \\
    \hline
\end{tabular}
\label{tab:results_EMU5TH_LCDM}
\end{table*}


%%%%%%%%%%%%%%%%%%%%%%%%%%%%%%%%%%%%%%%%%%%%%%%%%%

% Don't change these lines
\bsp	% typesetting comment
\label{lastpage}
\end{document}